\documentclass[aps,reprint,groupaddress,superscriptaddress,amssymb,prd]{revtex4-2}
\usepackage[utf8]{inputenc}
\usepackage{color}
\usepackage{xcolor}
\usepackage{graphicx}

\usepackage{mdframed}
\usepackage{physics}
\usepackage{amsmath}
\usepackage{amsfonts}
\usepackage{bm}
\usepackage[symbol]{footmisc}

\begin{document}
\preprint{APS/123-QED}

\title{Multiscale nonlinear integration drives accurate encoding of input information}

\author{Giorgio Nicoletti}
\affiliation{ECHO Laboratory, École Polytechnique Fédérale de Lausanne, Lausanne, Switzerland}
\author{Daniel Maria Busiello}
\affiliation{Max Planck Institute for the Physics of Complex Systems, Dresden, Germany}

\begin{abstract}
\noindent Biological and artificial systems encode information through several complex nonlinear operations, making their exact study a formidable challenge. These internal mechanisms often take place across multiple timescales and process external signals to enable functional output responses. In this work, we focus on two widely implemented paradigms: \textit{nonlinear summation}, where signals are first processed independently and then combined; and \textit{nonlinear integration}, where they are combined first and then processed. We study a general model where the input signal is propagated to an output unit through a processing layer via nonlinear activation functions. We demonstrate that integration systematically enhances input-output mutual information over a wide range of parameters and system sizes, while simultaneously enabling tunable input discrimination. Further, we reveal that high-dimensional embeddings and low-dimensional projections emerge naturally as optimal competing strategies. Our results uncover the foundational features of nonlinear information processing with profound implications for both biological and artificial systems.
\end{abstract}

\maketitle

\section{Introduction}
\noindent The ability to encode and process information from the external world is essential to maintain robust functioning in biological systems \cite{tkavcik2016information}. These goals are usually achieved through complex internal machinery that involves nonlinear operations. For example, multi-molecular reactions drive sensing and adaptation in chemical networks \cite{barkai1997robustness, aoki2019universal, flatt2023abc}, gene regulatory dynamics is controlled by protein-mediated interactions leading to multi-stable phases corresponding to different cell fates \cite{mochizuki2005analytical, trapnell2014dynamics}, phase coexistence phenomena sustain noise reduction and functional organization in cellular environments \cite{mitrea2016phase, klosin2020phase}, and complex interaction networks underlie the computational capabilities of neural populations \cite{vyas2020computation, barzon2022criticality, dubreuil2022role}.

As such, extracting information from a given input to generate a desired output is a fundamental problem that spans several fields, from signal processing in biochemical systems \cite{jordan2000signaling, cheong2011information} to designing and training artificial neural networks \cite{hagan1997neural, szandala2021review}. Many of these systems share the idea that inputs need to be processed via different types of nonlinear activation functions to enable non-trivial learning tasks. Despite remarkable results, understanding the key determinants of how the resulting architectures underpin information processing, and how the type of nonlinearity shapes its performances, is an active area of theoretical research \cite{karlik2011performance, ramachandran2017searching, apicella2021survey, nwankpa2018activation}. Recent works have investigated the performance of computation tasks instantiated by biological media, making an effort to bridge artificial and biochemical processing \cite{dack2024recurrent}. In particular, they have highlighted the pivotal computational role of nonlinear encoding \cite{chaudhuri2019impact, floyd2024limits, barzon2024maximal} and multiple timescales \cite{cavanagh2020diversity, golesorkhi2021brain, mariani2022disentangling, zeraati2024neural}.

Information theory provides us with tools to quantitatively study information-processing capabilities of various systems ranging from stochastic processes \cite{parrondo2015thermodynamics, ito2013information, nicoletti2021mutual, nicoletti2022mutual} to biological scenarios \cite{nicoletti2024informationgain, graf2024bifurcation, mattingly2021escherichia, bauer2023information}. While the impact of timescales on information propagation has been understood independently of the context \cite{nicoletti2024information}, the role of internal nonlinear mechanisms remains unclear without focusing on specific models. One of the main difficulties resides in the lack of general analytical approaches - only discrete-state systems that are sufficiently small can be embedded into an expanded state space that is tractable enough to solve the resulting master equation \cite{nicoletti2024information}. In fact, most general results obtained for large systems or phenomenological models rely on Gaussian approximations of various forms \cite{tostevin2009mutual, moor2023dynamic, nicoletti2024tuning}.

In this work, we overcome these limitations by analytically tackling a generic multiscale and dynamical information-processing system. A possibly high-dimensional signal is encoded by an input unit, processed by a processing unit, and finally passed on to an output unit. Interactions among the units form a general multilayer network structure that supports the propagation of the input information \cite{dedomenico2013mathematical, ghavasieh2020statistical, nicoletti2024information}. Crucially, each unit may operate on a different timescale and is composed of an arbitrary number of individual degrees of freedom (dofs), such as neurons in neural networks or chemical species in a signaling network. Crucially, operations between the units are implemented by different types of activation (or transfer) functions. We compare two paradigmatic processing schemes: \textit{nonlinear summation}, in which incoming signals are processed and then summed before affecting the target unit; and \textit{nonlinear integration}, in which the signal is first integrated, and then passed to the next unit through an activation function. We first find the exact expression for the probability distribution (pdf) of the system in different timescale regimes. Then, we employ this result to show that integration is associated with a higher mutual information between input and output units over a wide range of parameters and system sizes, emerging as the backbone of accurate processing. Further, we unravel a nontrivial interplay between the dimensionalities of the input and processing units, allowing us to define optimal operation regimes. Finally, we show that nonlinear integration also leads to the spontaneous emergence of bistability in the output layer even for Gaussian inputs, suggesting its crucial role in implementing input discrimination that can be tuned by tinkering with internal parameters.

\section{Results}
\subsection{Multiscale information-processing system}
\noindent To maintain the generality of the approach, we consider an information-processing system composed of three different stochastic units: input $I$, processing $P$, and output $O$. Each unit is composed of $M_\mu$ degrees of freedom with a shared timescale $\tau_\mu$, with $\mu = I, P, O$. All dofs within the same unit are linearly coupled with an interaction matrix $\hat{A}_{\mu}$ of size $M_\mu \times M_\mu$; conversely, the coupling from unit $\nu$ to $\mu$ is implemented via a nonlinear activation function $\vec{\phi}_{\mu\nu}$ that depends, in principle, on all dofs within $\nu$ and a $M_\mu \times M_\nu$ interaction matrix $\hat{A}_{\mu\nu}$. The system's dynamics is described by the following Langevin equations:
\begin{equation*}
    \tau_\mu \dot{\vec{x}}_\mu = -\hat{A}_{\mu} \vec{x}_\mu + \sum_{\nu \ne \mu} g_{\mu\nu} \vec{\phi}_{\mu\nu}(\hat{A}_{\mu\nu}; \vec{x}_\nu) + \sqrt{2\tau_\mu} \hat{\sigma}_\mu \vec{\xi}_\mu \;,
\end{equation*}
where $g_{\mu\nu}$ is an interaction strength between unit $\nu$ and $\mu$, $\hat{D}_{\mu} = \hat{\sigma}_\mu \hat{\sigma}_\mu^T$ a diagonal diffusion coefficient, and $\vec{\xi}_\mu$ a vector of Gaussian white noises.
The first observation is that, when all $g_{\mu\nu}$'s are zero, the units will converge to independent Gaussian distributions at stationarity. However, when interactions between units are turned on, information starts propagating across timescales. Furthermore, we may assume that the input evolves independently, as this is the case for a relevant class of biophysical scenarios \cite{nicoletti2024tuning,ma2009defining,rahi2017oscillatory,yi2000robust}. Then, the input is passed to the processing unit through a directional coupling, i.e., $g_{PI} \neq 0$ and $g_{IP} = 0$. After the processing step, the signal arrives at the output unit again through a directional coupling, i.e., $g_{OP} \neq 0$ and $g_{PO} = 0$. Importantly, our system can be easily extended to include several processing units, as our results are independent of their number.

To investigate how the mechanisms implementing internal nonlinear processing affect the information content of the system, we study the mutual information between input and output units,
\begin{eqnarray}
\label{eqn:mutual_info}
    I_{IO} &=& \int d\vec{x}_I d\vec{x}_O p_{IO}(\vec{x}_I, \vec{x}_O) \log_2\frac{p_{IO}(\vec{x}_I, \vec{x}_O)}{p_I(\vec{x}_I) p_O(\vec{x}_O)} = \\
    &=& H_O - \int d\vec{x}_I p_I(\vec{x}_I) h_{O|I}(\vec{x}_I) = H_O - \langle h_{O|I} \rangle_I \;, \nonumber
\end{eqnarray}
where $h_{O|I} = \langle p_{O|I} \log_2 p_{O|I} \rangle_O$. Here, $p_{IO}(\vec{x}_I, \vec{x}_O)$ is the joint pdf of input and output dofs, $p_I(\vec{x}_I)$ and $p_O(\vec{x}_O)$ are their respective marginal pdfs, and $H_O$ is the Shannon entropy \cite{cover1999elements} of the output unit computed in bits. $I_{IO}$ quantifies the information shared between $I$ and $O$, therefore acting as an unbiased proxy for processing accuracy in this paradigmatic setting \cite{cover1999elements}.

\begin{figure}
    \centering
    \includegraphics[width=\columnwidth]{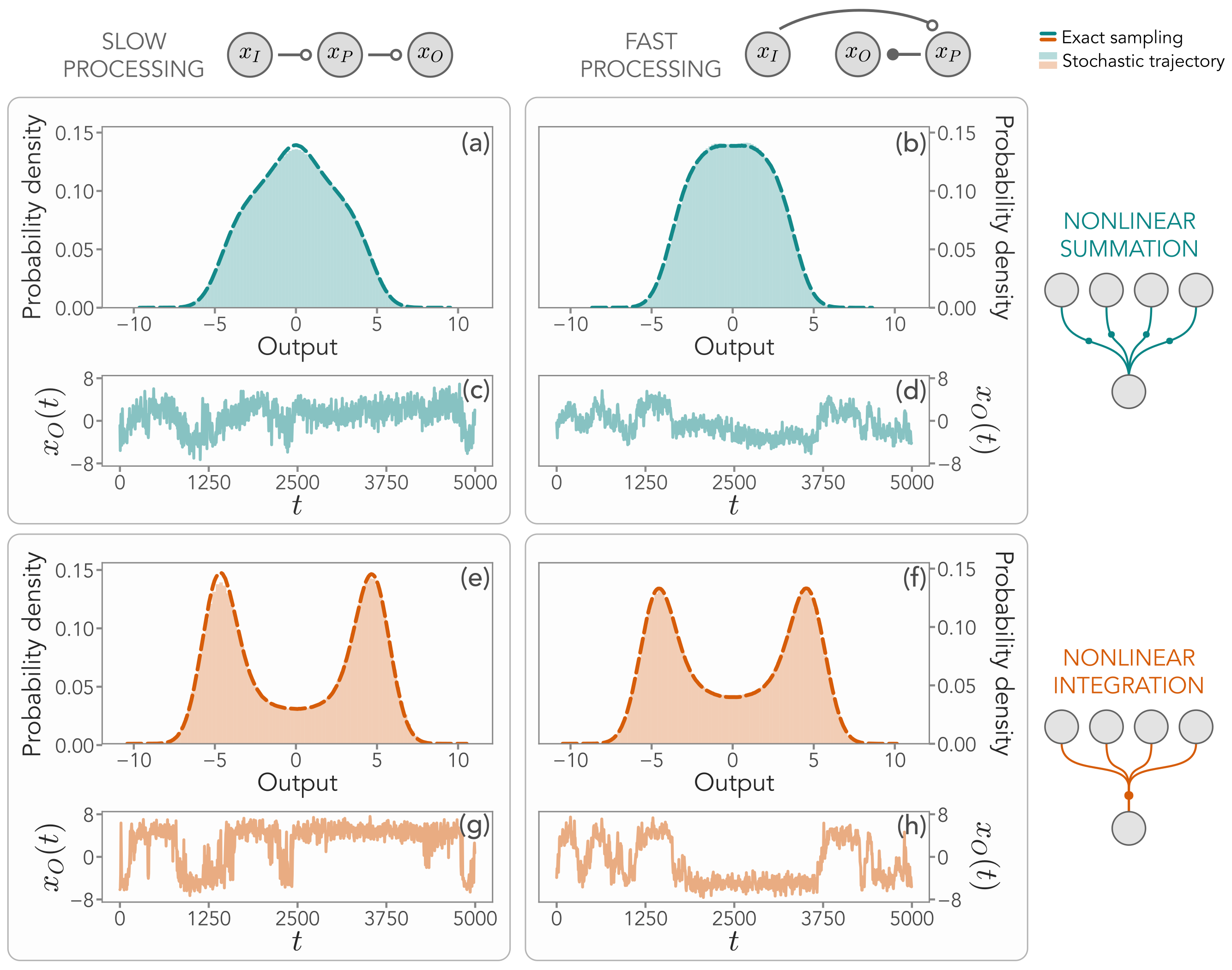}
    \caption{(a-d) Output distribution and stochastic trajectories of the output in the presence of nonlinear summation between units, both for a slow ($\tau_I \gg \tau_P \gg \tau_O$) and a fast ($\tau_I \gg \tau_O \gg \tau_P$) processing unit. In the model sketch, units are ordered from the slowest (left) to the fastest (right). Interactions are set to $g_{PI} = g_{OP} = 5$. (e-h) Same, but in the presence of nonlinear integration between units, where the activity of the components of each unit is first summed and then integrated. In panels (a-b) and (e-f), dashed lines are pdfs obtained by sampling the exact output distribution, and shaded areas the histograms obtained from Langevin trajectories with a timescale ratio $\Delta \tau = 10^{-2}$ between the units. In this figure, $M_I = 5$, $M_P = 3$, and $M_O = 1$. Interactions between units are distributed as $\mathcal{N}(0,1)$, whereas intra-unit interactions follow $\mathcal{N}(0, 0.9/\sqrt{M})$ to ensure input stability.}
    \label{fig:figure1}
\end{figure}

As demonstrated in \cite{nicoletti2024information}, if the dynamics of the input unit is the faster at play ($\tau_I \ll \tau_P, \tau_O$), no mutual information can be generated between $I$ and $O$. Conversely, a slow input is a necessary condition to have a non-zero $I_{IO}$. We still have the freedom to set the timescales of processing and output units, distinguishing two relevant cases: a fast-processing system ($\tau_P \ll \tau_O)$ and a slow-processing one ($\tau_P \gg \tau_O$). However, a crucial role is also played by the specific type of nonlinearity $\vec{\phi}_{\mu\nu}$ at hand. We distinguish two widely used but distinct cases, corresponding to different processing schemes: nonlinear summation ($\rm ns$) \cite{sompolinsky1988chaos, kadmon2015transition, engelken2023lyapunov} and integration ($\rm int$) \cite{lukosevicius2009reservoir, maheswaranathan2019universality, driscoll2024flexible, tanaka2016exploiting, malik2016multilayered}. They prescribe the following forms of interactions between units:
\begin{equation}
\begin{aligned}
\label{eqn:func_form}
(\phi^{i}_{\mu\nu})^{\rm ns} &= \frac{1}{M_\nu} \sum_{j=1}^{M_\nu} A_{\mu\nu}^{ij} \tanh(x_\nu^j) \\
(\phi^{i}_{\mu\nu})^{\rm int} &= \tanh\left(\frac{1}{M_\nu} \sum_{j=1}^{M_\nu} A_{\mu\nu}^{ij} x_\nu^j\right) \;,
\end{aligned}
\end{equation}
where all nodes in unit $\nu$ contribute to the dynamics of node $i$ in unit $\mu$ through a nonlinear activation function and a set of weights $A^{ij}_{\mu\nu}$, with $j = 1, \dots, M_\nu$, mediating the coupling. These two cases represent different physical processes. For a nonlinear summation, the signals generated by each dof in unit $\nu$ are first nonlinearly transformed, and then linearly projected by means of the interaction matrix $\hat{A}_{\mu\nu}$. In contrast, for a nonlinear integration, the signals from unit $\nu$ are first linearly combined via the weights $A^{ij}_{\mu\nu}$, and then the resulting integrated signal is nonlinearly transformed by the activation function and passed to the $i$-th dof of unit $\mu$. We set the activation function as a hyperbolic tangent to mimic customary modeling choices for neural networks \cite{sompolinsky1988chaos}. In Fig.~\ref{fig:figure1}, we show stochastic trajectories and pdfs of the output unit for slow and fast processing, both in the case of summation and integration. For simplicity of computation and visualization, we will consider a one-dimensional output unit throughout this manuscript. While there is no striking difference between slow and fast processing at the dynamical level, nonlinear summation and integration lead to two very different distributions in the output node. Integrating incoming signals from one unit to the other favors the spontaneous emergence of a pronounced switching behavior that reflects into a bistable distribution, a signature of input discrimination. The last part of this manuscript will be dedicated to quantitatively substantiating this observation.

\subsection{Exact solution for fast and slow processing units}
\noindent The first contribution of this Letter is to provide an analytical solution for the joint distribution of the whole system, $p_{IPO}$, that can be exploited to evaluate the input-output mutual information $I_{IO}$, and the output stationary pdf $p^{\rm st}_O$. While $I_{IO}$ informs us on the processing performance of the system, $p_O$ contains information on the ability to perform input discrimination. $p_{IPO}$ satisfies the following Fokker-Planck equation: 
\begin{equation*}
    \frac{\partial}{\partial t} p_{IPO} = \left( \frac{\mathcal{L}_I}{\tau_I} + \frac{\mathcal{L}_P}{\tau_P} + \frac{\mathcal{L}_O}{\tau_O} \right) p_{IPO}
\end{equation*}
where $\mathcal{L}_\mu$ is the Fokker-Planck operator associated with the unit $\mu = I, P, O$, as detailed in the Supplemental Material (SM) \cite{suppmat}. Although general exact expressions are out of reach without approximations, the limits of fast and slow processing can provide useful insights into system operations, provided the presence of a slow input unit. From \cite{nicoletti2024information, nicoletti2024gaussian}, we know that in these two limiting regimes the joint pdf of input, processing, and output units is the product of conditional distributions. As we show in the SM \cite{suppmat}, at stationarity we have:
\begin{eqnarray}
\label{eqn:p_fast}
    p^{\rm fp}_{IOP} &= p^{\rm st}_I p^{\rm st}_{P|I} p^{\rm eff, st}_{O|I} \qquad& {\rm fast ~processing} \\
\label{eqn:p_slow}
    p^{\rm sp}_{IOP} &= p^{\rm st}_I p^{\rm st}_{P|I} p^{\rm st}_{O|P} \qquad& {\rm slow ~processing} \;,
\end{eqnarray}
where the superscript ``$\rm st$" (omitted on the l.h.s.) stands for stationarity, and ``$\rm eff$" indicates a pdf that solves an effective operator obtained from the ensemble average over dofs faster than its corresponding unit. We use the superscript ``fp'' and ``sp'' to indicate that these quantities are evaluated respectively for fast and slow processing.
Let us inspect all these terms one by one. $p_I^{\rm st}$ is the multivariate Gaussian distribution of the input with mean $\vec{m}_I$ and covariance matrix $\hat{\Sigma}_I$ that solves the Lyapunov equation $\hat{A}_I \hat{\Sigma}_I + \hat{\Sigma}_I \hat{A}_I^T = 2 \hat{D}_I$. By exploiting the fact that intra-unit interactions are linear, all the conditional distributions may be written as:
\begin{equation}
\label{eqn:p_conditional}
    p^{\rm st}_{\mu|\nu} = \mathcal{N}\left(\vec{m}_{\mu|\nu}(\vec{x}_\nu), \hat{\Sigma}_\nu\right) \qquad \mu,\nu = I,P,O
\end{equation}
with $\hat{\Sigma}_\nu$ satisfying its corresponding Lyapunov equation, and the average containing the dependence on the conditional variable as follows:
\begin{equation}
    \label{eqn:average}
    \vec{m}_{\mu|\nu}(\vec{x}_\nu) = g_{\mu\nu} \hat{A}^{-1}_\mu \vec{\phi}_{\mu\nu}(\hat{A}_{\mu\nu}; \vec{x}_\nu)
\end{equation}
Notice that the functional form of Eq.~\eqref{eqn:average} depends on the nonlinear processing mechanism considered in Eq.~\eqref{eqn:func_form}. However, when an effective operator is involved, calculations become harder. By using a convergent expansion of the hyperbolic tangent, we show that:
\begin{eqnarray}
\label{eqn:p_conditional_effective}
    p^{\rm eff, st}_{O|I} = \mathcal{N}\left( \vec{m}^{\rm eff}_{O|I}(\vec{x}_I), \hat{\Sigma}_O \right)
\end{eqnarray}
with again $\hat{A}_O \hat{\Sigma}_O + \hat{\Sigma}_O \hat{A}_O^T = 2 \hat{D}_O$ and
\begin{equation}
\begin{aligned}
\label{eqn:average_eff}
    {\rm ns:} \quad \vec{m}_{O|I}^{\rm eff} &= g_{OP} \hat{A}_O^{-1} \left(\frac{\hat{A}_{OP}}{M_P} \vec{\mathcal{F}}(\vec{m}_{P|I},{\rm diag}(\hat{\Sigma}_P))\right) \\
    {\rm int:} \quad \vec{m}_{O|I}^{\rm eff} &= g_{OP} \hat{A}_O^{-1} \vec{\mathcal{F}}(\vec{m}_{\rm int},\vec{v}_{\rm int})
\end{aligned}
\end{equation}
where we employed the shorthand notation $\mathcal{F}^i(\vec{x},\vec{y}) = \mathcal{F}(x^i,y^i)$. In particular, $\mathcal{F}$ is a nontrivial nonlinear function defined in the SM \cite{suppmat}, and we introduced the following integrated quantities:
\begin{eqnarray*}
    \vec{m}_{\rm int} = \frac{1}{M_P} \hat{A}_{OP} \vec{m}_{P|I} \;, \qquad
    \vec{v}_{\rm int} = \frac{1}{M^2_P} \hat{A}_{OP} \hat{\Sigma}_P \hat{A}^T_{OP} \;.
\end{eqnarray*}
From Eq.~\eqref{eqn:average_eff}, we notice that the dependence on $\vec{x}_I$ enters solely through $\vec{m}_{O|P}$, defined in Eq.~\eqref{eqn:average}. The main difference resides in the fact that, in the case of summation, the nonlinear function $\vec{\mathcal{F}}$ has to be averaged with processing weights $\hat{A}_{OP}$, while in the case of integration, $\vec{\mathcal{F}}$ must be directly evaluated on integrated quantities. 

Putting all these results together, we obtain an analytical expression for the joint pdf of the whole system, $p_{IPO}$. We stress that $p_{IPO}$ is a highly nonlinear distribution. However, our factorization into conditional Gaussian distributions incorporates the nonlinearities only into their means, allowing in particular for efficient sampling. In Figs.~\ref{fig:figure1}a-b and \ref{fig:figure1}e-f, dashed curves indicate the output pdf obtained by sampling the exact joint distribution (details of the calculations are presented in the SM \cite{suppmat}). Furthermore, the structure of the resulting conditional dependencies is crucially different between fast and slow processing units, with fundamental implications for the mutual information between the input and the output.

\begin{figure*}
    \centering
    \includegraphics[width=\textwidth]{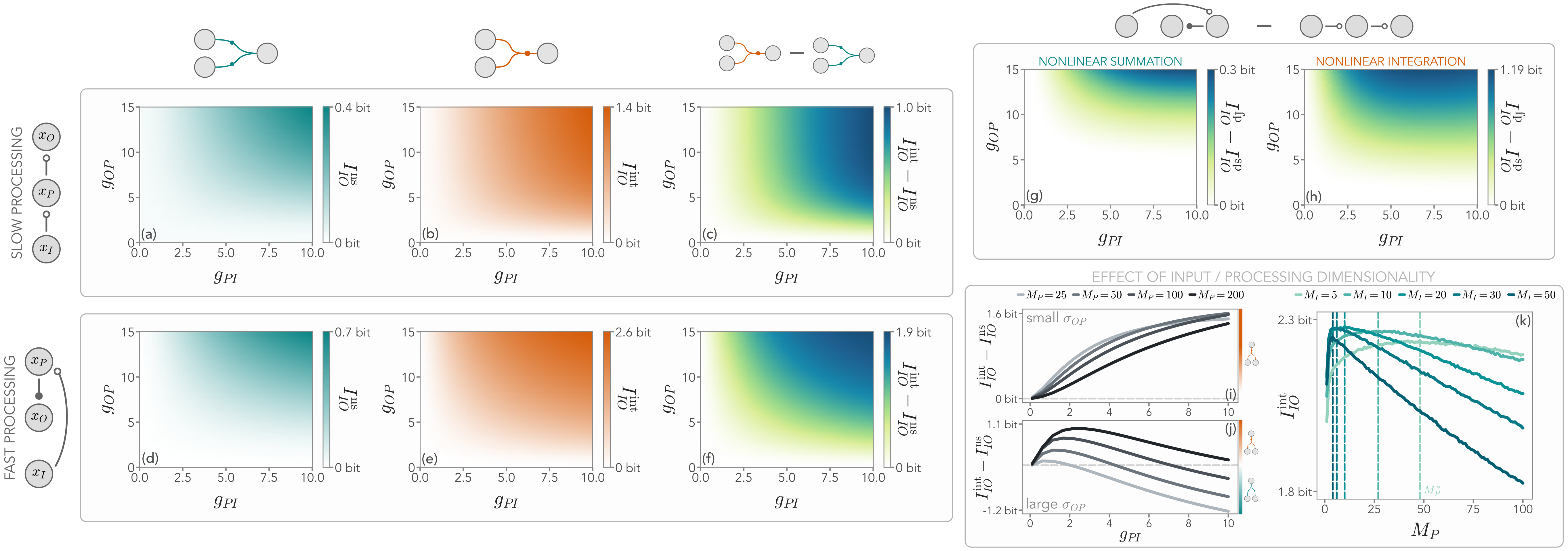}
    \caption{(a-c) Mutual information between input and output $I_{IO}$ in a system with a slow processing unit as a function of the coupling strengths $g_{PI}$ and $g_{OP}$. Nonlinear integration (superscript ``int'', orange) produces higher information than nonlinear summation (superscript ``ns'', teal). (d-f) Same, but for a fast processing unit. In the model sketch, the units are ordered from the fastest (top or right) to the slowest (bottom or left). (g-h) Furthermore, $I_{IO}$ is systematically higher with a fast processing unit (Eq.~\eqref{eqn:mutual_OI_fast}). This effect is particularly relevant for an activation function implementing nonlinear integration. In these panels, $\sigma_{PI} = \sigma_{OP} = 1$, $\sigma_{I} = \sigma_{P} = 0.9$, and $M_I = M_P = 50$. In this figure, results are obtained by averaging over $10^3$ realization of the random interaction matrices. (i-k) Here, we focus on the fast processing scenario. (i-j) The advantage of nonlinear integration depends on the interplay between the dimensionality of the processing unit, $M_P$, and the variance of $\hat{A}_{OP}$, $\sigma_{OP}^2$. For small variances ($\sigma_{OP} = 0.1$), we always find $I_{IO}^\mathrm{int} > I_{IO}^\mathrm{ns}$. However, at large variances ($\sigma_{OP} = 10$), such advantage is achieved only for large enough $M_P$. (k) In the nonlinear integration case, for a given input dimension $M_I$ there exists an optimal processing dimensionality $M_P^*$ that maximizes the input-output information (dashed lines). As $M_I$ increases, $M_P$ decreases.}
    \label{fig:figure2}
\end{figure*}

\subsection{Enhanced information by nonlinear integration}
\noindent We can now exploit the exact factorization of the joint pdf of the system to evaluate the accuracy of processing the stochastic input and encoding it into the one-dimensional output, by means of the mutual information $I_{IO}$ in Eq.~\eqref{eqn:mutual_info}. In the case of fast processing, the joint distribution of input and output is obtained from Eq.~\eqref{eqn:p_fast} by integrating over $\vec{x}_P$, i.e., $p^{\rm fp}_{IO} = p^{\rm st}_I p^{\rm eff, st}_{O|I}$. Thus, since $p^{\rm eff, st}_{O|I}$ is a Gaussian distribution with a variance independent of $\vec{x}_I$ (Eq.~\eqref{eqn:p_conditional_effective}), $h^{\rm fp}_{O|I}$ does not depend on $\vec{x}_I$ and is equal to:
\begin{equation*}
    h_{O|I}^{\rm fp} = \frac{1}{2} \left[ M_O (1 + \log_2 (2 \pi) ) + \log_2 \det(\hat{\Sigma}_O) \right]
\end{equation*}
so that the mutual information reads:
\begin{equation}
\label{eqn:mutual_OI_fast}
    I_{IO}^{\rm fp} = H^{\rm fp}_O - h_{O|I}^{\rm fp} \;.
\end{equation}
To evaluate this quantity is sufficient to compute the output pdf and evaluate its Shannon entropy using standard estimators \cite{vasicek1976test,kozachenko1987sample} (see SM \cite{suppmat} for details).
In the presence of a slow processing unit, instead, from Eq.~\eqref{eqn:p_slow} we have:
\begin{equation}
\label{eqn:p_IO_slow}
    p^{\rm sp}_{IO} = p^{\rm st}_I \int d\vec{x}_P p^{\rm st}_{O|P} p^{\rm st}_{P|I} = p^{\rm st}_I p^{\rm st}_{O|I} \;.
\end{equation}
Although an expression for $h_{O|I}^{\rm sp}$ cannot be easily obtained, we can efficiently sample $p^{\rm st}_{O|I}$ by using Eq.~\eqref{eqn:p_IO_slow}, as detailed in the SM \cite{suppmat}, to compute the mutual information $I_{IO}^{\rm sp}$.

In Fig.~\ref{fig:figure2}, for both slow and fast processing, we compare the mutual information between input and output for the case of nonlinear summation, $I_{IO}^{\rm ns}$, and integration, $I_{IO}^{\rm int}$. We omit the superscripts referring to the processing timescale whenever it is already clear from the picture. Interactions within the same unit $\mu$ are distributed as $\mathcal{N}(0,\sigma_\mu/\sqrt{M_\mu})$, with $\sigma_\mu < 1$ to ensure stability without inter-unit couplings. The dependence of the mutual information on input stability is studied in the SM \cite{suppmat} and resembles the findings of \cite{nicoletti2024gaussian}. Interaction between units $\mu$ and $\nu$ are distributed as $\mathcal{N}(0,\sigma_{\mu\nu})$. All results are obtained by averaging over realizations of these random matrices. Figs.~\ref{fig:figure2}a-c and \ref{fig:figure2}d-f respectively show that, independently of the internal timescale ordering, nonlinear integration always leads to higher mutual information with respect to summation. Interestingly, this discrepancy seems to be systematically more pronounced in the presence of a fast processing unit (see Fig.~\ref{fig:figure2}g-h). This fundamental result suggests that integration, possibly coupled with proper timescale ordering, may serve as a foundational scheme to support accurate processing in both biological and artificial systems.

Additionally, the computational advantage of nonlinear integration holds for any processing dimensions and coupling strength when interactions between the processing and output units have small variance, $\sigma^2_{OP}$. For larger $\sigma^2_{OP}$, increasing the dimensionality of the processing unit appears beneficial for processing accuracy (see Fig.~\eqref{fig:figure2}i-j), while at small $M_P$ we find a limited strong-coupling regime where nonlinear summation provides larger mutual information. Intuitively, this happens since, for small processing sizes, the elements of $\hat{A}_{OP}$ are not evenly sampled from their Gaussian distribution. This effect becomes less and less prominent, and eventually disappears, as $M_P$ increases. Furthermore, by inspecting the interplay between input and processing dimensionality, we find that at a given $M_I$, there exists an optimal value of $M_P$ that maximizes $I_{IO}^{\rm int}$, as shown in Fig.~\ref{fig:figure2}k (see also SM \cite{suppmat}). Remarkably, the optimal processing size decreases by increasing the input size, suggesting that a nonlinear embedding of a low-dimensional input in a higher-dimensional processing space favors information encoding. On the contrary, $I_{IO}^{\rm int}$ is maximal at small $M_P$ for large $M_I$, so that information processing is favored by a nonlinear compression of the input in a lower-dimensional processing space. This behavior reveals quantitative insights into optimal operation regimes and diverse strategies to encode information.

\subsection{Emergent output bistability}
\begin{figure}
    \centering
    \includegraphics[width=\columnwidth]{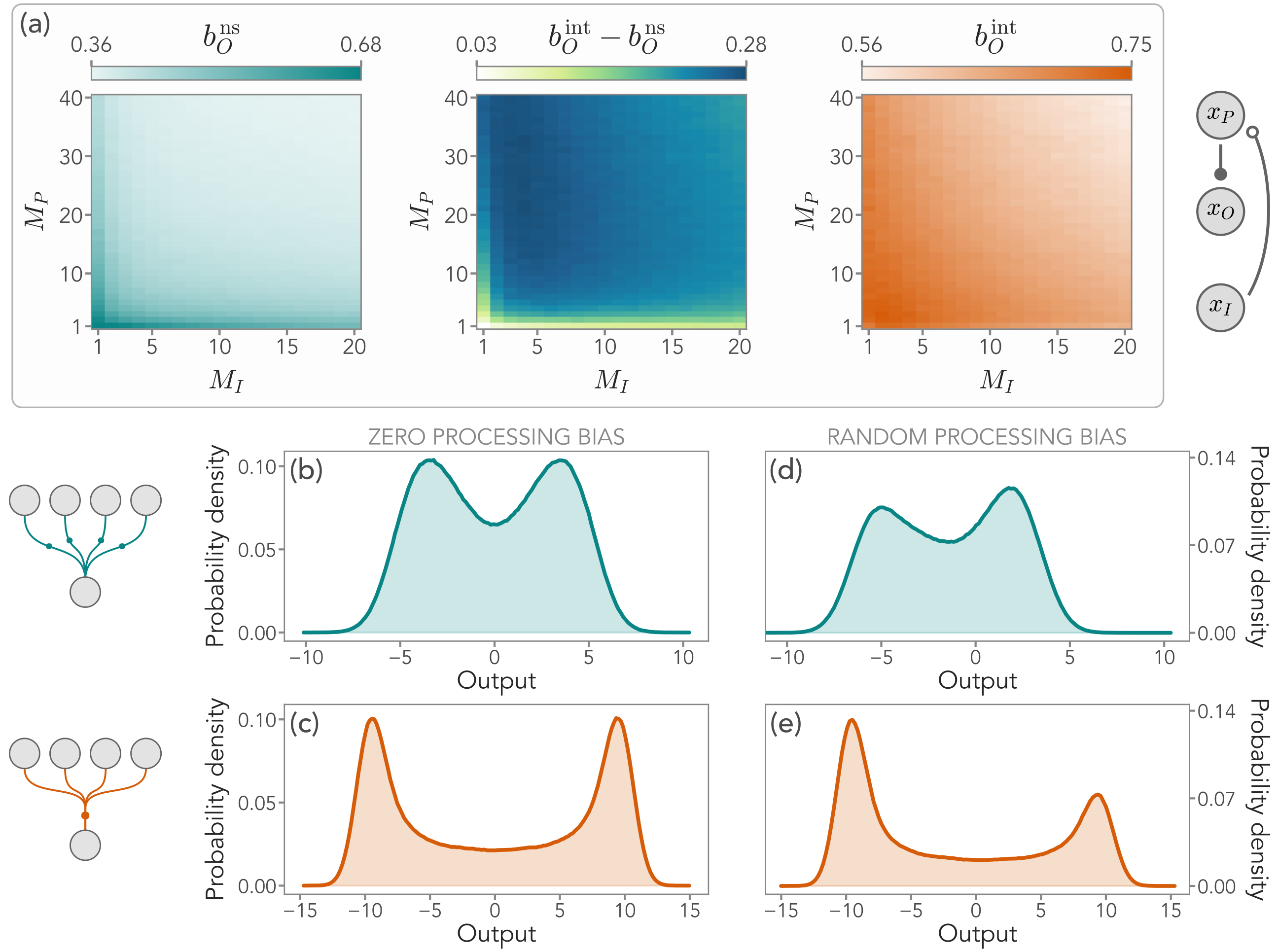}
    \caption{(a) Bimodality coefficient of output pdf, $b_O$, as a function of input and processing dimensions in a system with a fast processing unit. Here, $g_{PI} = g_{OP} = 10$, $\sigma_{OP} = \sigma_{PI} = 1$, and $\sigma_{I} = \sigma_{P} = 0.9$. Results are averaged over $10^3$ realization of random matrices. Both activation functions, implementing nonlinear summation (``ns'', teal) or nonlinear integration (``int'', orange), may lead to a bistable output distribution, particularly for smaller dimensions. Notwithstanding, on average nonlinear integration enhances bimodality in all parameter ranges explored. (b-c) Output distributions for $M_I = 5$ and $M_P = 10$. (d-e) Same random matrix realization as in panels (b-c), but after introducing a random bias in the processing nonlinearity. The bias allows for tuning the emergent output bistability, allowing the system to select one of the modes. Importantly, due to its randomness, the bias may have opposite effects depending on the type of activation function.}
    \label{fig:figure3}
\end{figure}

\noindent Nonlinear integration is also advantageous from a dynamical perspective. In Fig.~\ref{fig:figure3}a, we consider the case of a fast processing unit and compare the bistability of the output pdf obtained for nonlinear summation and integration by means of the Staple's bimodality coefficient (see SM \cite{suppmat}), respectively denoted by $b_O^{\rm ns}$ and $b_O^{\rm int}$. We show that integration is associated with higher bistable coefficients, i.e., more pronounced bistability, thus enabling more accurate input discrimination in the output distribution. The presence of a stochastic Gaussian input and random interaction matrices makes it particularly hard to pinpoint the input features that the system is discriminating. However, the robustness of our finding for such a general scenario hints at an intrinsic advantage of integration in favoring input discrimination. 

Crucially, however, this emergent bistability may be tuned by introducing suitable processing biases in how the signal of certain nodes is encoded. We can add this ingredient in Eq.~\eqref{eqn:func_form} by applying the substitution $x_\nu^j \to x_\nu^j - \theta_\nu^j$, where the bias is introduced as:
\begin{equation}
\begin{aligned}
\label{eqn:func_form_bias}
(\phi^{i}_{\mu\nu})^{\rm ns, b} &= \frac{1}{M_\nu} \sum_{j=1}^{M_\nu} A_{\mu\nu}^{ij} \tanh(x_\nu^j - \theta^j_\nu) \\
(\phi^{i}_{\mu\nu})^{\rm int, b} &= \tanh\left(\frac{1}{M_\nu} \sum_{j=1}^{M_\nu} A_{\mu\nu}^{ij} (x_\nu^j - \theta_\nu^j)\right) \;,
\end{aligned}
\end{equation}
where the superscript ``b'' indicates the presence of the bias. Note that
$\vec{\theta}_\nu$ can be in principle different for each unit. In Fig.~\eqref{fig:figure3}b-e, we consider a system in the presence of a fast processing unit and include the presence of a random bias $\vec{\theta}_\nu$ whose elements are drawn from $\mathcal{N}(0,1)$ for each $\nu$. By comparing the same realization of random interaction matrices $\hat{A}_\mu$ and $\hat{A}_{\mu\nu}$ (as discussed for Fig.~\ref{fig:figure2}) without (Fig.~\ref{fig:figure3}b-c) and with (Fig.~\ref{fig:figure3}d-e) $\vec{\theta}_\nu$, we find that the presence of a bias triggers an unbalance in the output bistability. Notice that this emergent unbalance can be different between summation and integration due to the intrinsic randomness of all system's components, as shown in Fig.~\ref{fig:figure3}. Thus, nonlinear integration enables more pronounced tunable output bistability, allowing the system to statistically select one of the two emerging modes.

\section{Discussion}
\noindent In this work, we studied nonlinear processing through different activation functions in a paradigmatic three-unit system. By leveraging the presence of multiple timescales, we analytically obtained the joint distribution of the system and computed the input-output mutual information. We compared two nonlinear processing schemes, summation and integration, employed in multiple contexts. In systems implementing nonlinear summation, inputs are first processed and then averaged, while inputs are first averaged and then processed in systems supporting nonlinear integration. We showed that integration generally leads to higher input-output mutual information, allowing for increased processing accuracy. At the same time, this internal scheme enables tunable input discrimination. Finally, we highlighted a nontrivial competing behavior between high-dimensional embedding and low-dimensional projection in the processing space, depending on the input dimensionality. Overall, our paradigmatic approach allowed us to quantitatively investigate the emergence of accurate encoding in information-processing architectures encompassing key ingredients common to both biological and artificial systems. In doing so, we highlighted the unforeseen advantages of nonlinear integration under very general conditions, uncovering its role in shaping information in multiscale systems. Our results might inspire the design of efficient artificial processing schemes and the discovery of the principles guiding biological computation.

Extensions are manifold. Future works may systematically investigate other types of activation functions, evaluating their performances in terms of input-output information and the relative timescales between the units. Furthermore, it will be interesting to consider systems where different units implement different activation functions, allowing for more heterogeneity in terms of computational capabilities. Along this line, architectures with several processing units, possibly acting on a diverse range of timescales, may be necessary to deal with bio-inspired models and more structured inputs. This setting will also enable a natural implementation of the existence of multiple tasks whose presence might dramatically change the definition of processing performance. Our work will stand as a foundational step for these explorations, unraveling how different types of dynamical nonlinearities underlie information and computation in real-world systems.

\begin{acknowledgments}
\noindent G.N. acknowledges funding provided by the Swiss National Science Foundation through its Grant CRSII5\_186422. The authors acknowledge the support of the Munich Institute for Astro-, Particle and BioPhysics (MIAPbP), funded by the Deutsche Forschungsgemeinschaft (DFG, German Research Foundation) under Germany´s Excellence Strategy – EXC-2094 – 390783311, where this work was first conceived during the MOLINFO workshop.
\end{acknowledgments}

\pagebreak

\newpage
\widetext
\pagebreak

\setcounter{equation}{0}
\setcounter{figure}{0}
\setcounter{table}{0}
\setcounter{page}{1}
\setcounter{section}{0}
\setcounter{subsection}{0}
\makeatletter
\renewcommand{\theequation}{S\arabic{equation}}
\renewcommand{\thefigure}{S\arabic{figure}}
\renewcommand{\thesection}{S\Roman{section}} 


\begin{center}\large{\bf Supplemental Material: ``Multiscale nonlinear integration drives accurate encoding of input information''}\end{center}

\section{Multilayer model with nonlinear interactions between units}
\noindent We consider a stochastic system whose degrees of freedom (dofs) can be partitioned into $N$ components (or units for now on), each containing $M_\mu$ variables that evolve with a shared timescale $\tau_\mu$, for $\mu = 1, \dots, N$. These partitions can be viewed as different units or layers in a multilayer network, so that the $i$-th node in the $\mu$-th layer describes a continuous stochastic variable $x_\mu^i$, with $i = 1, \dots, M_\mu$. Thus, the dimensionality of unit $\mu$ corresponds to the number of nodes $M_\mu$ in the corresponding layer, with nodes representing its interacting internal degrees of freedom. We will often refer to $x_\mu^i$ as the activity of the node, in reminiscence of models for the dynamics of neural networks, even if this framework is amenable to describe a vast variety of biological and artificial systems, as discussed in the main text. The multilayer network is described by an adjacency tensor $A_{\mu\nu}^{ij}$, which measures the strength of the interaction going from the node $x_\nu^j$ to the node $x_\mu^i$. Hence, the matrix $\hat{A}_{\mu\mu}$ describes the interactions between the nodes of the $\mu$-th unit, and $\hat{A}_{\mu\nu}$ describes the interactions between unit $\nu$ and $\mu$. For stability, we take the self-interactions to be $A_{\mu\mu}^{ii} = 1$.

In full generality, we can describe the dynamical evolution of the system by the set of Langevin equations
\begin{equation}
\label{eqn:SM:langevin}
    \tau_\mu \dot{x}_\mu^i = -\sum_{j=1}^{M_\mu}A_{\mu\mu}^{ij}f_\mu\left(x_\mu^j\right) + \sum_{\nu \ne \mu} g_{\mu\nu}\phi_{\mu\nu}\left(A_{\mu\nu}^{i, 1}, \dots, A_{\mu\nu}^{i, M_\nu}; x_\nu^1, \dots, x_\nu^{M_\nu}\right) + \sqrt{2 D_\mu^i \tau_\mu}\xi_\mu^i
\end{equation}
where $f_\mu$ is a generic activation function characterizing the intra-unit interactions between nodes, $\phi_{\mu\nu}$ is another activation function dictating the inter-unit interactions from nodes in unit $\nu$ to the node $x_\mu^i$, $g_{\mu\nu}$ is the corresponding interaction strength, $D_\mu^i$ a constant noise strength, and $\xi_\mu^i$ are independent white noises. In the main text, we introduced $\vec{\phi}_{\mu\nu}$, defined as $\vec{\phi}_{\mu\nu}(\hat{A}_{\mu\nu}; \vec{x}_\nu) \equiv \phi(A_{\mu\nu}^{i, 1}, \dots, A_{\mu\nu}^{i, M_\nu}; x_\nu^1, \dots, x_\nu^{M_\nu})$. In other words, the dependence on the node is reflected in the dependence on matrix elements and components of $\vec{x}_\nu$, while the functional form of the activation function stays unchanged. In this work, we set $f_\mu(z) = z$, so that, in the absence of inter-unit interactions, the local dynamics of a unit corresponds to an analytically solvable Gaussian process. Notice also that, in this picture, inter-unit interactions can link the dynamics of multiple nodes belonging to different units and, therefore, evolving on different timescales.

We focus on two different kinds of activation functions between the units. In the first case, we consider the coupling between the units to be implemented as a nonlinear summation, i.e.,
\begin{equation}
\label{eqn:SM:NS}
    \phi_{\mu\nu}^\mathrm{ns}\left(A_{\mu\nu}^{i, 1}, \dots, A_{\mu\nu}^{i, M_\nu}; x_\nu^1, \dots, x_\nu^{M_\nu}\right) = \frac{1}{M_\nu} \sum_{j = 1}^{M_\nu} A_{\mu\nu}^{ij}\tanh{x_\nu^j}
\end{equation}
where $\tanh$ plays the role of a nonlinear activation function. In this case, the activities of the nodes of the $\nu$-th unit are first nonlinearly transformed and then linearly averaged to obtain the overall interaction with $x_\mu^i$. In the second case, instead, we consider an activation function implementing a nonlinear integration of the whole unit activity, i.e.,
\begin{equation}
\label{eqn:SM:INT}
    \phi_{\mu\nu}^\mathrm{int}\left(A_{\mu\nu}^{i, 1}, \dots, A_{\mu\nu}^{i, M_\nu}; x_\nu^1, \dots, x_\nu^{M_\nu}\right) = \tanh\left(\frac{1}{M_\nu} \sum_{j = 1}^{M_\nu} A_{\mu\nu}^{ij} x_\nu^j\right) \; .
\end{equation}
In Eq.~\eqref{eqn:SM:INT}, the overall activity of unit $\nu$, defined as the average activity weighted on the corresponding interactions with $x_\mu^i$, is \emph{nonlinearly integrated}, i.e., it enters in the activation function as a whole. These two choices have been employed in several contexts, particularly in reservoir computing \cite{lukosevicius2009reservoir} and more in general in random recurrent neural networks \cite{sompolinsky1988chaos, kadmon2015transition, engelken2023lyapunov, maheswaranathan2019universality, driscoll2024flexible}. However, they lead to qualitatively different dynamics and deeply affect dependencies between the units.

In general, solving Eq.~\eqref{eqn:SM:langevin} amounts to solving the corresponding Fokker-Planck equation
\begin{equation}
\label{eqn:SM:Fokker-Planck}
    \pdv{}{t}p_{1, \dots, N}(\vec{x}_1, \dots, \vec{x}_N, t) = \sum_{\mu = 1}^N \, \frac{1}{\tau_\mu}\mathcal{L}_\mu p_{1, \dots, N}(\vec{x}_1, \dots, \vec{x}_N, t)
\end{equation}
where $\vec{x}_\mu = (x_\mu^1, \dots, x_\mu^{M_\mu})$, $p_{1, \dots, N}(\vec{x}_1, \dots, \vec{x}_N, t)$ is the multilayer joint probability - with each layer representing a specific unit $\mu$ - describing the probability of the activity of all nodes at time $t$, and $\mathcal{L}_\mu$ is the Fokker-Planck operator of unit $\mu$:
\begin{equation}
    \mathcal{L}_\mu = \sum_{i=1}^{M_\mu}\pdv{}{x_\mu^i} \left[\sum_{j=1}^{M_\mu}A_{\mu\mu}^{ij}x_\mu^j - \sum_{\nu \ne \mu} g_{\mu\nu} \phi_{\mu\nu}\left(A_{\mu\nu}^{i, 1}, \dots, A_{\mu\nu}^{i, M_\nu}; x_\nu^1, \dots, x_\nu^{M_\nu}\right) + D_\mu^i \pdv{}{x_\mu^i}\right] \; .
\end{equation}
Eq.~\eqref{eqn:SM:Fokker-Planck} is a highly nonlinear equation in $\vec{x}_\mu$ and thus solving it exactly is a formidably challenging task. However, we will exploit the fact that interactions within a unit are linear to obtain an analytically tractable factorization of the nonlinear joint probability $p_{1, \dots, N}$ in a timescale-separation limit.

\section{Input-output systems and mutual information}
\noindent A relevant class of systems (see main text for references) is characterized by the presence of an input unit $I$ which evolves independently on the rest of the system - so that $A_{I \nu} = 0$ for all $\nu$ - and an output unit $O$, whose nodes are not a source of any interactions to other units - $A_{\nu O} = 0$ for all $\nu$. These systems exhibit a hierarchical structure allowing for an clear identification of the signal to be read (the input) and the variables that encode it (the output). In particular, we are interested in computing the mutual information between the input and the output, namely
\begin{align}
    I_{IO} & = \int d\vec{x}_I d \vec{x}_O \, p_{IO}(\vec{x}_I, \vec{x}_O) \log_2 \frac{p_{IO}(\vec{x}_I, \vec{x}_O)}{p_{I}(\vec{x}_I) p_O(\vec{x}_O)} \nonumber \\
    & = H_O - H_{O|I}
    \label{IIO}
\end{align}
where $H_O$ is the differential entropy of the output and $H_{O|I}$ the conditional entropy,
\begin{equation*}
    H_O = -\int d\vec{x}_O \, p_O(\vec{x}_O) \log_2 p_O(\vec{x}_O), \qquad H_{O|I} = -\int d\vec{x}_I d \vec{x}_O \, p_{IO}(\vec{x}_I, \vec{x}_O) \log_2 p_{O|I}(\vec{x}_O | \vec{x}_I) \; .
\end{equation*}
The mutual information quantifies the dependencies between the output and the input units in terms of how much information they share, Specifically, it captures the reduction of the uncertainty in the output, quantified by its entropy, once the input is known. Notice that the expression in Eq.~\eqref{IIO} holds for any system even without a hierarchical structure, however the identification of inputs and outputs might become more complex, eventually leading to ambiguity in the interpretation of the results.

For what follows, it will be useful to rewrite the mutual information in terms of the function
\begin{equation}
    h_{O|I}(\vec{x}_I) = -\int d\vec{x}_O \, p_{O|I}(\vec{x}_O | \vec{x}_I) \log_2 p_{O|I}(\vec{x}_O | \vec{x}_I)
\end{equation}
which is nothing but the entropy of the conditional distribution $p_{O|I}$, and whose expectation value over the input distribution is exactly the conditional entropy. Thus, we have 
\begin{align}
\label{eqn:SM:mutual}
    I_{IO} = H_O - \ev{h_{O|I}}_I = H_O - \int d\vec{x}_I \, p_I(\vec{x}_I) h_{O|I}(\vec{x}_I)
\end{align}
which, as we will see, will allow us to evaluate the mutual information directly from samples of the joint distribution.

\section{Direct input-output connections}
\label{sec:SM:IO}
\noindent Within the hierarchical scheme highlighted above, we first consider the instructive example of a system with only two units before considering the full three-unit system of the main text: an input unit, $\vec{x}_I$, with $M_I$ nodes, and an output unit, $\vec{x}_O$, with $M_O$ nodes. The adjacency tensor is given by
\begin{equation}
    \hat{A} = \begin{pmatrix}
        \hat{A}_I & 0 \\
        \hat{A}_{OI} & \hat{A}_O
    \end{pmatrix}
\end{equation}
where $\hat{A}_I$ and $\hat{A}_O$ are $M_I \times M_I$ and $M_O \times M_O$ matrices, respectively, and $\hat{A}_{OI}$ describes the connections from the input to the output unit. This model contains two timescales and any processing mechanism converting input into output is effectively taken into account by the nonlinear inter-unit activation functions, as presented above. In order to obtain analytical solutions for the joint probability $p_{IO}(\vec{x}_I, \vec{x}_O, t)$, we focus on the limiting case of a slow input, $\tau_I \gg \tau_O$. Indeed, the other limit, albeit interesting from a mathematical perspective, will not give rise to any information between the units, as proven in \cite{nicoletti2024information}. We also focus on the steady-state solution, so we will neglect time dependencies.

In this limit we can rescale time by the slowest timescale, i.e., $t \to t/\tau_I$, and seek a solution of the form
\begin{equation*}
    p_{IO}(\vec{x}_I, \vec{x}_O, t) = p_{IO}^{(0)}(\vec{x}_I, \vec{x}_O, t) + \frac{\tau_O}{\tau_I} p_{IO}^{(1)}(\vec{x}_I, \vec{x}_O, t) + \mathcal{O}\left(\left(\frac{\tau_O}{\tau_I}\right)^2\right)
\end{equation*}
where the superscript denotes the order of the expansion in the small parameter, $\tau_O/\tau_I \ll 1$. Up to the first order, i.e., with $\mathcal{O}(\tau_O/\tau_I)$ corrections, the Fokker-Planck equation becomes
\begin{equation*}
    \pdv{}{t}p_{IO}^{(0)} = \frac{\tau_I}{\tau_O}\mathcal{L}_O p_{IO}^{(0)} + \mathcal{L}_I p_{IO}^{(0)} + \mathcal{L}_O p_{IO}^{(1)} + \mathcal{O}\left(\frac{\tau_O}{\tau_I}\right)
\end{equation*}
where we suppressed the dependencies of $p_{IO}$ for brevity. Proceeding order-by-order, as outlined in \cite{nicoletti2024gaussian}, we obtain a solution for $p_{IO}^{(0)} := p_{IO}^\mathrm{st}$ of the form
\begin{equation}
\label{eqn:SM:pIO_twoL}
    p_{IO}^\mathrm{st}(\vec{x}_I, \vec{x}_O) = p_I^\mathrm{st}(\vec{x}_I) \, p_{O|I}^\mathrm{st}(\vec{x}_O | \vec{x}_I)
\end{equation}
where the probabilities are the stationary solutions of the operators
\begin{equation}
    \mathcal{L}_I(\vec{x}_I) \, p_I^\mathrm{st}(\vec{x}_I) = 0, \qquad \mathcal{L}_{O|I}(\vec{x}_I, \vec{x}_O) \, p_{O|I}^\mathrm{st}(\vec{x}_O | \vec{x}_I) = 0
\end{equation}
with
\begin{equation}
    \mathcal{L}_{O|I}(\vec{x}_I, \vec{x}_O) = \sum_{i=1}^{M_O}\pdv{}{x_O^i} \left[\sum_{j=1}^{M_O}A_{O}^{ij}\left(x_O^j - \sum_{k=1}^{M_O} g_{OI}\left(A_O^{-1}\right)^{jk}\phi_{OI}\left(A_{OI}^{k, 1}, \dots, A_{OI}^{k, M_I}; \vec{x}_I\right)\right) + D_O^i \pdv{}{x_O^i}\right] \; .
\end{equation}
Crucially, both these operators admit a stationary multivariate Gaussian distribution $\mathcal{N}(\vec{m}, \hat{\Sigma})$, where $\vec{m}$ is the mean and $\hat{\Sigma}$ the covariance matrix. Indeed, the input evolves independently of the output and, as such, it is governed by a Fokker-Planck equation without inter-unit interactions. On the other hand, the conditional probability of the output given the input, $p^{\rm st}_{O|I}$, evolves according to an operator where interactions between the units only depend on the value of $\vec{x}_I$, which has to be considered quenched. Specifically, we have that
\begin{equation}
    p_I(\vec{x}_I)^\mathrm{st} = \mathcal{N}\left(\vec{0}, \hat{\Sigma}_I\right), \qquad p_{O|I}^\mathrm{st}(\vec{x}_O | \vec{x}_I) = \mathcal{N}\left(\vec{m}_{O|I}(\vec{x}_I), \hat{\Sigma}_O\right)
\end{equation}
where 
\begin{equation}
    m^i_{O|I}(\vec{x}_I) = g_{OI}\sum_{k=1}^{M_O}\left(A_O^{-1}\right)^{ik}\phi_{OI}\left(A_{OI}^{k, 1}, \dots, A_{OI}^{k, M_I}; \vec{x}_I\right), \quad i = 1, \dots, M_O
\end{equation}
and the covariance matrices obey the Lyapunov equations $\hat{A}_I \hat{\Sigma}_I + \hat\Sigma_I\hat{A}_I^T = 2 \hat{D}_I$, with $\hat{D}_I = \mathrm{diag}\left(D_I^1, \dots, D_I^{M_I}\right)$, and similarly for the output unit. The fundamental advantage of the factorization in Eq.~\eqref{eqn:SM:pIO_twoL} is that both its components are Gaussian. We also note that the structure of the solution remains identical independently on the form of the nonlinearity, so it is valid for both Eq.~\eqref{eqn:SM:NS} and Eq.~\eqref{eqn:SM:INT}.

Yet, since the mean of $p_{O|I}^\mathrm{st}$ explicitly depends on $\vec{x}_I$, the output distribution, $p^{\rm st}_O = \int d\vec{x}_I ~p^{\rm st}_{O|I} ~p^{\rm st}_{I}$, obtained by marginalizing over $\vec{x}_I$, will not be Gaussian due to the nonlinearity of $\phi_{OI}$. Therefore, the output entropy appearing in the mutual information in Eq.~\eqref{eqn:SM:mutual} cannot be evaluated analytically. However, the second term is given by
\begin{equation}
    h_{O|I}(\vec{x}_I) = \frac{1}{2}\left[M_O \left(1 + \log_2 (2\pi)\right) + \log_2 \det \hat\Sigma_O\right] \equiv h_{O|I}
\end{equation}
which does not depend on $\vec{x}_I$. As a consequence, we just need to evaluate $H_O$ numerically. In particular, we can easily sample the joint distribution $p_{IO}^\mathrm{st}$, and thus obtain samples from the output that can be used to estimate its distribution and entropy. We employ both a classic Vasicek estimator \cite{vasicek1976test} and the standard Kozachenko-Leonenko kNN estimator \cite{kozachenko1987sample}, as we focus on the case of one-dimensional outputs. Due to the curse of dimensionality, in higher dimensions more refined strategies would be needed \cite{lu2020enhancing}. With an estimate $\tilde{H}_O$ of the output entropy at hand, the mutual information between the output and input unit simply reads
\begin{equation}
    I_{IO} = \tilde{H}_O -\frac{1}{2}\left[M_O \left(1 + \log_2 (2\pi)\right) + \log_2 \det \hat\Sigma_O\right] \; .
\end{equation}
Although the joint probability in Eq.~\eqref{eqn:SM:pIO_twoL} is nonlinear, the factorization into Gaussian conditional probabilities allows for an efficient sampling scheme. Indeed, we can obtain a sample $\{\vec{x}_I, \vec{x}_O\}$ by sampling first the Gaussian input, and then the output from $p_{O|I}^\mathrm{st}$, whose mean depend on $\vec{x}_I$. 

\subsection{Effect of the linear stability of the input}
\begin{figure}
    \centering
    \includegraphics[width=0.9\textwidth]{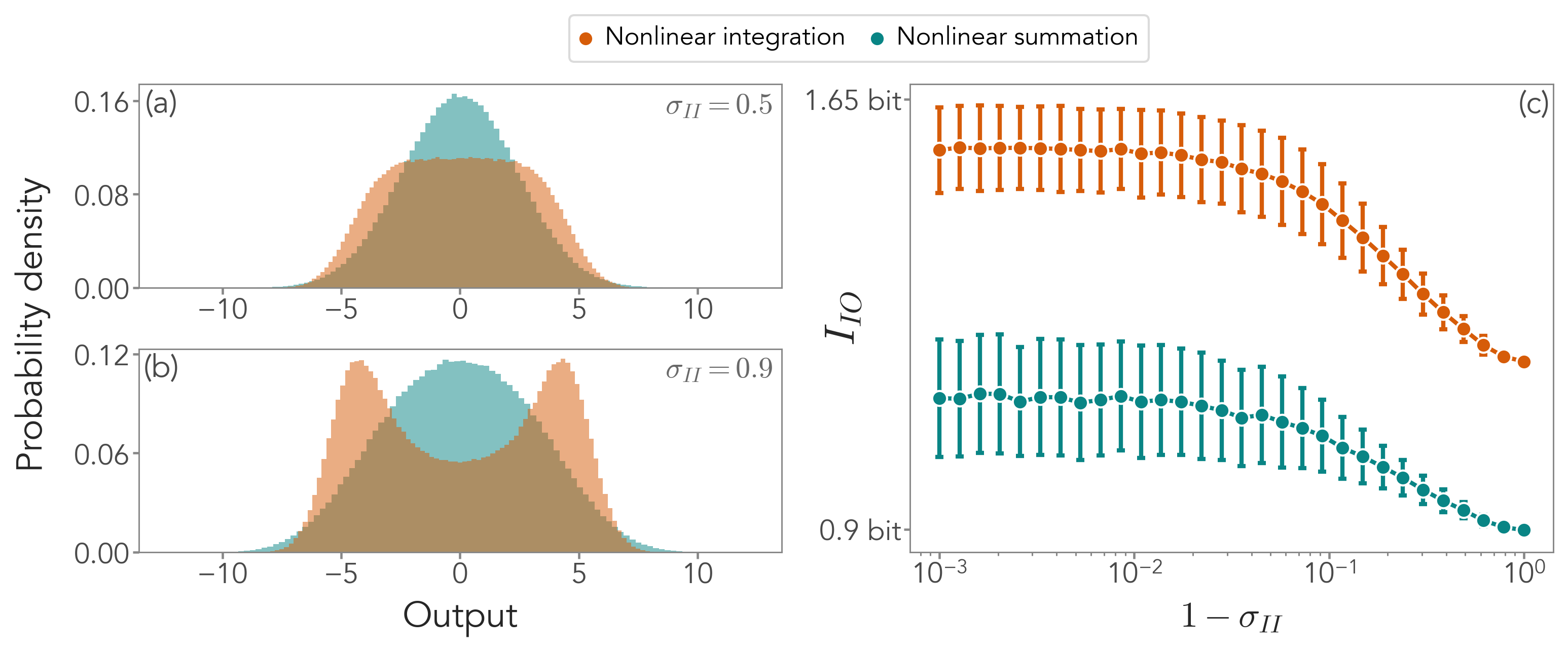}
    \caption{Effect of the input stability in an input-output system, both with an activation function following a nonlinear summation and a nonlinear integration. In this figure, $M_I = 50$, $M_O = 1$, $g_{OI} = 5$, $\sigma_{OI} = 5$. (a-b) Output distributions for two different values of $\sigma_{II}$. As the input is closer to the edge of stability $\sigma_{II}^c = 1$, the distribution with nonlinear integration becomes bistable. (c) The mutual information between the input and the output, $I_{IO}$, increases as the input approaches the edge of linear stability, regardless of the type of activation function at hand. The increase is more marked for the nonlinear integration case. }
    \label{fig:SM_input_stability}
\end{figure}

\noindent We now study how the linear stability of the input affects the two-unit system described in the previous section. The input is stable if the spectral radius of $\hat{A}_I$, i.e., the absolute value of its largest eigenvalue, is smaller than $A_{\mu\mu}^{ii} = 1$. We consider an input-output system whose connections are described by the random matrices
\begin{equation}
    A_I^{ij} \sim \mathcal{N} \left(0, \frac{\sigma_{II}}{\sqrt{M_I}}\right), \quad A_{OI}^{ij} \sim \mathcal{N}\left(0, \sigma_{OI}\right)
\end{equation}
where the usual normalization of the standard deviation of the elements of $\hat{A}_I$ ensures a proper scaling of its spectrum with the input dimensionality. Similarly,
\begin{equation}
    A_{OI}^{ij} \sim \mathcal{N}\left(0, \sigma_{OI}\right)
\end{equation}
and we fix $\sigma_{OI} = 1$ for simplicity. In Figure \ref{fig:SM_input_stability} we show that $\sigma_{II}$ has a crucial impact on the output probability and the mutual information between the input and the output. In particular, $p_O$ shows an emergent bistability as $\sigma_{II}$ increases (Figure \ref{fig:SM_input_stability}a-b). Furthermore, as expected from previous works \cite{barzon2024maximal}, $I_{IO}$ increases as $\sigma_{II}$ approaches the edge of linear stability $\sigma_{II}^c$ (Figure \ref{fig:SM_input_stability}c). The increase is more marked for the case of the nonlinear integration case, suggesting that, as the input becomes more variable in time, an activation function with nonlinear integration is able to better track the input features.

\section{Intermediate processing between the input and the output}
\noindent Since it is known that timescales play a pivotal role in shaping information propagation \cite{nicoletti2024information}, we will explore their interplay with nonlinear interactions by adding an intermediate processing unit to the system that relays information. Since processing can, in principle, act on a different timescale, this more general system is characterized by three units: an input unit $\vec{x}_I$, with $M_I$ nodes; a processing unit $\vec{x}_P$, with $M_P$ nodes; and an output unit $\vec{x}_O$, with $M_O$ nodes. The adjacency tensor is
\begin{equation}
    \hat{A} = \begin{pmatrix}
        \hat{A}_I & 0 & 0 \\
        \hat{A}_{PI} & \hat{A}_P & 0\\
        0 & \hat{A}_{OP} & \hat{A}_O
    \end{pmatrix}
\end{equation}
so that the input is given to the processing unit before it is passed on to the output. To retain a clear interpretation of input and output signals, we also maintain the hierarchical structure of the model. As before, $\hat{A}_I$, $\hat{A}_P$ $\hat{A}_O$ are $M_I \times M_I$, $M_P \times M_P$ and $M_O \times M_O$ matrices, respectively. $\hat{A}_{PI}$ describes the connections from the input to the processing unit, and $\hat{A}_{OP}$ the connections from the processing to the output unit. In order to study this system exactly, following \cite{nicoletti2024information}, we again consider the input to be the slowest dof at play - so that it can generate a non-zero mutual information - with timescale $\tau_I$. Then, we distinguish between two cases: slow processing, i.e., $\tau_I \gg \tau_P \gg \tau_O$, and fast processing, i.e., $\tau_I \gg \tau_O \gg \tau_P$. As in the previous section, we aim to compute the mutual information between the input and the output, which we take to be one-dimensional for numerical stability.

\subsection{Fast processing}
\noindent We proceed as in the two-unit case. After rescaling time as $t \to t/\tau_I$, we seek a solution of the form
\begin{equation*}
    p_{IPO}(\vec{x}_I, \vec{x}_P, \vec{x}_O, t) = p_{IPO}^{(P, 0)}(\vec{x}_I, \vec{x}_P, \vec{x}_O, t) + \epsilon_P\left[p_{IPO}^{(O, 0)}(\vec{x}_I, \vec{x}_P, \vec{x}_O, t) + \epsilon_O p_{IPO}^{(I, 1)}(\vec{x}_I, \vec{x}_P, \vec{x}_O, t) \right]  + \mathcal{O}\left(\epsilon_P^2, \epsilon_O^2\right)
\end{equation*}
where $\epsilon_O = \tau_O/\tau_I \gg \epsilon_P = \tau_P/\tau_I$, and the superscript denotes the order in the corresponding term, distinguishing also between the variable it refers to, since there are as many zeroth order as the number of variables faster than the input. At leading order, the Fokker-Planck equation becomes
\begin{equation*}
    \pdv{}{t}p_{IPO}^{(P, 0)} = \left[\mathcal{L}_I + \frac{\mathcal{L}_O}{\epsilon_O} + \frac{\mathcal{L}_P}{\epsilon_P}\right]p_{IPO}^{(P, 0)} + \mathcal{L}_P p_{IPO}^{(O, 0)}
\end{equation*}
which can be solved order-by-order. At order $\mathcal{O}(1/\epsilon_P)$, we find that
\begin{equation*}
    \mathcal{L}_P (\vec{x}_I, \vec{x}_P) \, p_{IPO}^{(P, 0)}(\vec{x}_I, \vec{x}_P, \vec{x}_O, t) = 0
\end{equation*}
and, after a marginalization over the output state, $\vec{x}_O$, we have:
\begin{equation*}
    \mathcal{L}_P (\vec{x}_I, \vec{x}_P) \, p_{IP}^{(P, 0)}(\vec{x}_I, \vec{x}_P, t) = 0 = p_I^{(P,0)} (\vec{x}_I, t) \mathcal{L}_P (\vec{x}_I, \vec{x}_P) \, p_{P|I}^\mathrm{st}(\vec{x}_P | \vec{x}_I) \implies \mathcal{L}_{P|I} (\vec{x}_I, \vec{x}_P) \, p_{P|I}^\mathrm{st}(\vec{x}_P | \vec{x}_I) = 0
\end{equation*}
where we introduced $\mathcal{L}_{P| I} := \mathcal{L}_P$, to emphasize that its stationary distribution $p_{P|I}^\mathrm{st}$ is evaluated at fixed input. Thus, a solution of the form $p_{IPO}^{(P, 0)} = p_{P|I}^\mathrm{st} p_{IO}^{(P, 0)}$ automatically solves the first order. In particular, as in the previous section, we have that
\begin{equation}
    p_{P|I}^\mathrm{st}(\vec{x}_P | \vec{x}_I) = \mathcal{N}\left(\vec{m}_{P | I} (\vec{x}_I), \hat{\Sigma}_P\right)
\end{equation}
is a Gaussian distribution with a mean that depends nonlinearly on the input,
\begin{equation}
    m^i_{P|I}(\vec{x}_I) = g_{PI} \sum_{k=1}^{M_P}\left(A_P^{-1}\right)^{ik}\phi_{PI}\left(A_{PI}^{k, 1}, \dots, A_{PI}^{k, M_I}; \vec{x}_I\right), \quad i = 1, \dots, M_P,
\end{equation}
and a covariance obeying the Lyapunov equation $\hat{A}_P \hat{\Sigma}_P + \hat\Sigma_P\hat{A}_P^T = 2 \hat{D}_P$.

At the next order, $\mathcal{O}(1/\epsilon_O)$, we have that
\begin{equation*}
    p_{P|I}^\mathrm{st}(\vec{x}_P | \vec{x}_I) \mathcal{L}_O(\vec{x}_P, \vec{x}_O) \, p_{IO}^{(P, 0)}(\vec{x}_I, \vec{x}_O, t) = 0 = p_{P|I}^\mathrm{st}(\vec{x}_P | \vec{x}_I) p_I^{(P,0)} (\vec{x}_I, t)  \mathcal{L}_O(\vec{x}_P, \vec{x}_O) \, p_{O|I}^{(P, 0), \mathrm{st}}(\vec{x}_O | \vec{x}_I) \; .
\end{equation*}
To solve for $p_{O|I}^{(P, 0), \mathrm{st}}$, due to the explicit dependence of $\mathcal{L}_O(\vec{x}_P, \vec{x}_O)$ on the processing state, we integrate over $\vec{x}_P$ to obtain the following effective Fokker-Planck operator
\begin{equation}
\label{eqn:SM:mPP_effective_operator_definition}
    \mathcal{L}_O^\mathrm{eff} (\vec{x}_I, \vec{x}_O) := \int d\vec{x}_P \, p_{P|I}^\mathrm{st}(\vec{x}_P | \vec{x}_I) \mathcal{L}_O(\vec{x}_P, \vec{x}_O)
\end{equation}
leading to:
\begin{equation*}
    p_I^{(P,0)} (\vec{x}_I, t) \mathcal{L}_O^\mathrm{eff} (\vec{x}_I, \vec{x}_O) \, p_{O|I}^{(P, 0), \mathrm{st}}(\vec{x}_O | \vec{x}_I) = 0 \; .
\end{equation*}
Thus, if we introduce the effective stationary distribution
\begin{equation}
    \mathcal{L}_O^\mathrm{eff} (\vec{x}_I, \vec{x}_O) p_{O|I}^\mathrm{eff, st}(\vec{x}_O | \vec{x}_I) = 0 ,
\end{equation}
we end up with $p_{IPO}^{(P, 0)} = p_{P|I}^\mathrm{st} p_{O|I}^\mathrm{eff, st}p_I^{(P, 0)}$. However, to find an explicit form for $p_{O|I}^\mathrm{eff, st}$, we need to find an analytic expression for the $\mathcal{L}_O^\mathrm{eff}$, and solve for its stationary state. By construction, we have:
\begin{align*}
    \mathcal{L}_O^\mathrm{eff} (\vec{x}_I, \vec{x}_O) & = \sum_{i=1}^{M_O}\pdv{}{x_O^i} \left[\sum_{j=1}^{M_O}A_{O}^{ij}\left(x_O^j - g_{OP}\sum_{k=1}^{M_O}\left(A_O^{-1}\right)^{jk}\ev{\phi_{OP}^k}_P(\vec{x}_I)\right) + D_O^i \pdv{}{x_O^i}\right]
\end{align*}
with
\begin{equation*}
    \ev{\phi_{OP}^k}_P(\vec{x}_I) = \frac{1}{\sqrt{(2\pi)^{M_P} \det \hat\Sigma_P}}\int d\vec{x}_P \exp\left[-\frac{1}{2}(\vec{x}_P - \vec{m}_{P|I}(\vec{x}_I))^T\hat\Sigma_P^{-1}(\vec{x}_P - \vec{m}_{P|I}(\vec{x}_I))\right] \phi_{OP}\left(\vec{A}_{OP}^k, \vec{x}_P\right)
\end{equation*}
where we introduced $\vec{A}_{OP}^k = (A_{OP}^{k, 1}, \dots, A_{OP}^{k, M_P})$ for brevity. 

\subsubsection{Nonlinear summation}
\begin{figure}
    \centering
    \includegraphics[width=0.9\textwidth]{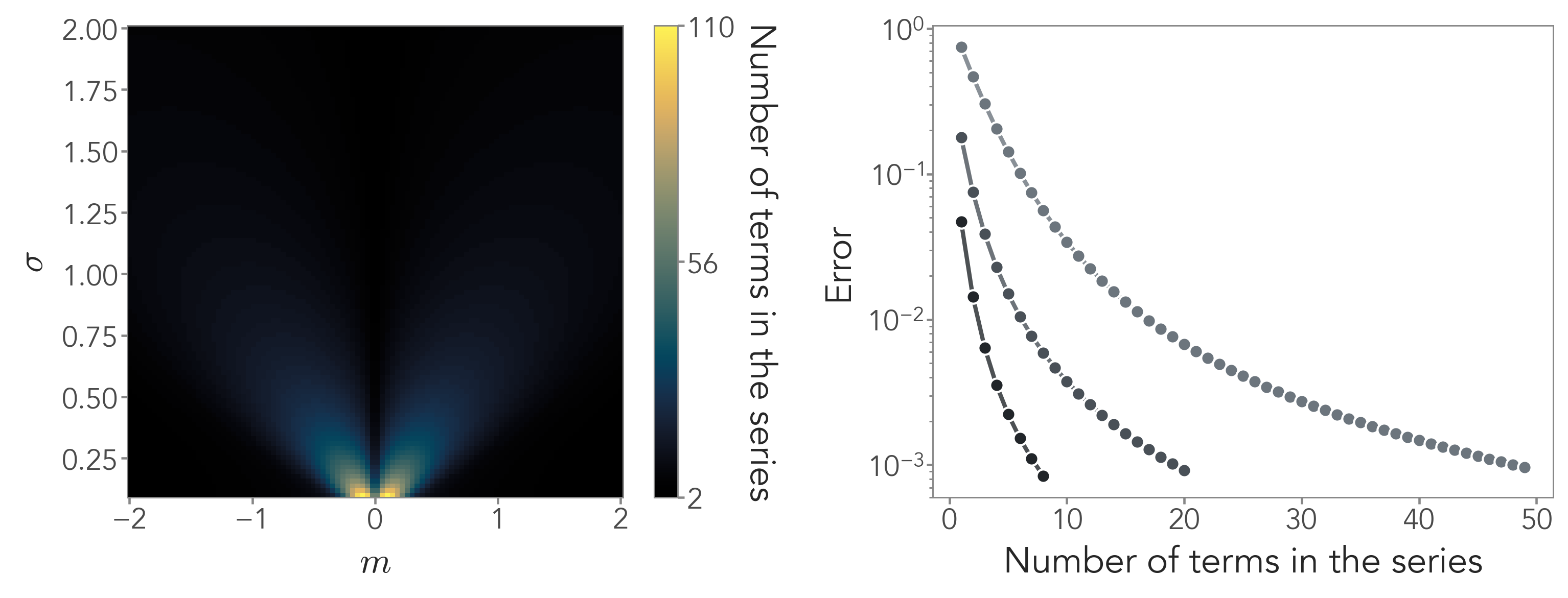}
    \caption{Convergence of the series in Eq.~\eqref{eqn:SM:apptanh} with respect to the numerical integration of $\ev{\tanh(z)}_G$. The contour plot shows, for different mean and standard deviation, the number of terms in the series needed to reach an error of $10^{-3}$, i.e., an absolute difference between the truncated series and the numerical integral difference of $10^{-3}$.}
    \label{fig:SM:tanh_series}
\end{figure}

We now need to distinguish the two classes of nonlinear couplings in Eqs.~\eqref{eqn:SM:NS} and \eqref{eqn:SM:INT}, as the integral over $\vec{x}_P$ explicitly depends on the form employed for activation function implementing interactions between the units. We first consider the case of a nonlinear summation (indicating it by with the superscript ns), Eq.~\eqref{eqn:SM:NS}, so that
\begin{align*}
    \ev{\phi_{OP}^i}_P^\mathrm{ns}(\vec{x}_I) & \propto \int d x_P^1 \dots d x_P^j \dots d x_P^{M_P} \exp\left[-\frac{1}{2}(\vec{x}_P - \vec{m}_{P|I}(\vec{x}_I))^T\hat\Sigma_P^{-1}(\vec{x}_P - \vec{m}_{P|I}(\vec{x}_I))\right]  \sum_{j = 1}^{M_P} \frac{A_{OP}^{ij}}{M_P}\tanh{x_P^j} \\
    & \propto \sum_{j = 1}^{M_P} \frac{A_{OP}^{ij}}{M_P} \int d x_P^j \exp\left[-\frac{1}{2 \Sigma_P^{jj}} \left(x_P^j - m_{P|I}^j(\vec{x}_I)\right)^2\right] \tanh x_P^j
\end{align*}
where we neglected the normalization terms for brevity and exploited the fact that the hyperbolic tangent term only depends on the components of $\vec{x}_P$ separately, so that the marginalization of the Gaussian over the other $M_P - 1$ components is straightforward and amounts to eliminating their corresponding rows and columns of the covariance matrix. However, finding an exact expression of the Gaussian average of the hyperbolic task is not trivial. To proceed further, we need first to rewrite the hyperbolic tangent as
\begin{align*}
    \tanh{z} & = \frac{\sinh{z}}{\cosh{z}} = \frac{1 - e^{-2z}}{1 + e^{-2z}} = (1 - e^{-2z}) \sum_{n = 0}^{\infty} (-1)^n e^{-2nz} \\
    & = \sum_{n = 0}^{\infty} (-1)^n \left[e^{-2nz} - e^{-2(n+1)z}\right] = \sum_{n = 0}^{\infty} (-1)^n e^{-2nz} + \sum_{n = 1}^\infty (-1)^n e^{-2nz} \\
    & = 1 + 2 \sum_{n = 1}^\infty (-1)^n e^{-2nz}
\end{align*}
which is convergent if and only if $e^{-2z} < 1$, i.e., for $z > 0$. We now need to evaluate its Gaussian average, here also indicated as $\ev{\tanh(z)}_G$, namely
\begin{align*}
    \frac{1}{\sqrt{2 \pi \sigma^2}}\int_{-\infty}^{+\infty} dz \tanh(z) e^{-\frac{(z-m)^2}{2\sigma^2}} & = \frac{1}{\sqrt{2 \pi \sigma^2}}\left[\int_{0}^{+\infty} dz \tanh(z) e^{-\frac{(z-m)^2}{2\sigma^2}} + \int_{-\infty}^0 dz \tanh(z) e^{-\frac{(z-m)^2}{2\sigma^2}}\right] \\
    & = \frac{1}{\sqrt{2 \pi \sigma^2}}\left[\int_{0}^{+\infty} dz \tanh(z) e^{-\frac{(z-m)^2}{2\sigma^2}} - \int_{0}^{+\infty} dz \tanh(z) e^{-\frac{(z+m)^2}{2\sigma^2}}\right] \\
    & = \frac{1}{\sqrt{2 \pi \sigma^2}}\left[\int_{0}^{+\infty} dz \left(1 + 2 \sum_{n = 1}^\infty (-1)^n e^{-2nz}\right) e^{-\frac{(z-m)^2}{2\sigma^2}} - e^{-\frac{(z+m)^2}{2\sigma^2}}\right] \\
    & = \erf\left(\frac{m}{\sqrt{2\sigma^2}}\right) + 2 \sum_{n = 1}^\infty (-1)^n \int_{0}^{+\infty} \frac{dz}{\sqrt{2 \pi \sigma^2}} e^{-2nz} \left[e^{-\frac{(z-m)^2}{2\sigma^2}} - e^{-\frac{(z+m)^2}{2\sigma^2}}\right]
\end{align*}
which is a convergent expression since the integral is evaluated for $z > 0$. Thus,
\begin{align*}
    \ev{\tanh(z)}_G & = \erf\left(\frac{m}{\sqrt{2\sigma^2}}\right) + \sum_{n = 1}^\infty (-1)^n \left[e^{2n (n\sigma^2 - m)}\left(\erf\left(\frac{m - 2n\sigma^2}{\sqrt{2\sigma^2}}\right) + 1\right) - e^{2n (n\sigma^2 + m)}\left(\erf\left(\frac{-m - 2n\sigma^2}{\sqrt{2\sigma^2}}\right) + 1\right)\right] \\
    & = \erf\left(\frac{m}{\sqrt{2\sigma^2}}\right) + \sum_{n = 1}^\infty (-1)^n e^{2n^2\sigma^2} \left[e^{- 2 n m}\left(1 + \erf\left(\frac{m - 2n\sigma^2}{\sqrt{2\sigma^2}}\right)\right) - e^{2n m}\mathrm{erfc}\left(\frac{m + 2n\sigma^2}{\sqrt{2\sigma^2}}\right)\right]
\end{align*}
where $\mathrm{erfc}(z) = 1 - \erf(z)$. It is convenient to rewrite this expression as
\begin{equation}
\label{eqn:SM:apptanh}
    \ev{\tanh(z)}_G = \erf\left(\frac{m}{\sqrt{2\sigma^2}}\right) + \sum_{n = 1}^\infty (-1)^n e^{2 n^2 \sigma^2} \left[V_n^+(m, \sigma^2) - V_n^-(m, \sigma^2)\right]
\end{equation}
where we introduced the functions
\begin{equation}
    \begin{gathered}
        V_n^+(m, \sigma^2) = e^{-2nm} \mathrm{erfc}\left(\frac{2n\sigma^2 - m}{\sqrt{2\sigma^2}}\right) \\
        V_n^-(m, \sigma^2) = e^{2nm} \mathrm{erfc}\left(\frac{2n\sigma^2 + m}{\sqrt{2\sigma^2}}\right)
    \end{gathered} \; .
\end{equation}
Eq.~\eqref{eqn:SM:apptanh} is an exact expression for the Gaussian average of the hyperbolic tangent. In practice, it can be efficiently evaluated numerically, as convergence with the number of terms in the sum over $n$ is typically quick, depending on the value of the mean and the variance, as shown in Figure \ref{fig:SM:tanh_series}.

Therefore, we immediately have that
\begin{align}
    \mathcal{L}_O^\mathrm{ns,eff} (\vec{x}_I, \vec{x}_O) & = \sum_{i=1}^{M_O}\pdv{}{x_O^i} \left[\sum_{j=1}^{M_O}A_{O}^{ij}\left(x_O^j - m_{O|I}^{\mathrm{ns}, j}(\vec{x}_I)\right) + D_O^i \pdv{}{x_O^i}\right]
\end{align}
where the mean is given by
\begin{align}
\label{eqn:SM:mOgI_FP_NS}
    m_{O|I}^{\mathrm{ns}, i}(\vec{x}_I) = \frac{g_{OP}}{M_P} \sum_{j=1}^{M_O} \sum_{k = 1}^{M_P}\left(A_O^{-1}\right)^{ij}A_{OP}^{jk} \Biggl[& \erf\left(\frac{m_{P|I}^k(\vec{x}_I)}{\sqrt{2\Sigma_P^{kk}}}\right) + \nonumber \\
    & + \sum_{n = 1}^\infty (-1)^n e^{2 n^2 \Sigma_P^{kk}} \left[V_n^+\left(m_{P|I}^k(\vec{x}_I), \Sigma_P^{kk}\right) - V_n^-(m_{P|I}^k(\vec{x}_I), \Sigma_P^{kk})\right] \Biggr] \; .
\end{align}
By comparison with the formulas of the main text, $\mathcal{F}(x^k,v^k)$ is equal to the term in the square brackets with $x^k \to m_{P|I}^k$ and $v^k \to \Sigma^{kk}_P$ in this case. The same will hold for all other similar expressions presented herein. We included the superscript NS also in the effective operator to emphasize the case here considered. Notice that the mean of the output inherits a direct dependence on the input $\vec{x}_I$ through the averaging procedure over the fast processing, a phenomenon typical of minimal propagation paths in timescale-separated multiscale systems. For a detailed discussion beyond the scope of this work, see \cite{nicoletti2024information}.

\subsubsection{Nonlinear integration}
\noindent We now switch to the case of nonlinear integration, Eq~\eqref{eqn:SM:INT}. Contrary to the previous case, we cannot reduce the effective operator to a sum of independent one-dimensional integrals. Indeed, we have
\begin{align*}
    \mathcal{L}_O^\mathrm{int, eff} (\vec{x}_I, \vec{x}_O) & = \sum_{i=1}^{M_O}\pdv{}{x_O^i} \left[\sum_{j=1}^{M_O}A_{O}^{ij}\left(x_O^j - g_{OP}\sum_{k=1}^{M_O}\left(A_O^{-1}\right)^{jk}\ev{\phi_{OP}^k}^\mathrm{int}_P(\vec{x}_I)\right) + D_O^i \pdv{}{x_O^i}\right]
\end{align*}
where
\begin{align*}
    \ev{\phi_{OP}^i}^\mathrm{int}_P(\vec{x}_I) & \propto \int d x_P^1 \dots d x_P^{M_P} \exp\left[-\frac{1}{2}(\vec{x}_P - \vec{m}_{P|I}(\vec{x}_I))^T\hat\Sigma_P^{-1}(\vec{x}_P - \vec{m}_{P|I}(\vec{x}_I))\right] \tanh\left(\sum_{j = 1}^{M_P} \frac{A_{OP}^{ij}}{M_P}{x_P^j}\right) \; .
\end{align*}
Here, we are using the superscript INT to highlight that these quantities only refer to the nonlinear integration case. In order to exploit the expansion of the hyperbolic tangent, we consider the change of variables $\vec{z} = \hat{C}_i \vec{x}_P$, where
\begin{equation}
    \label{eqn:changevariable}
    \hat{C}_i = \begin{pmatrix}
        \mathbb{I}_{M_P - 1} & \vec{0} \\
        \vec{A}_{OP}^{i}/M_P
    \end{pmatrix}
\end{equation}
is a $M_P \times M_P$ matrix and $\vec{A}_{OP}^{i} = ({A}_{OP}^{i1}, \dots, {A}_{OP}^{iM_P})$. Thus, $\vec{z} = (z^1, \dots, z^{M_P}) = (x_P^1, \dots, x_P^{M_P - 1}, y_i)$ with $y_i = 1/M_P \sum_j A_{OP}^{ij}x_P^j$. After applying the change of variables, we end up with
\begin{align*}
    \ev{\phi_{OP}^i}^\mathrm{int}_P(\vec{x}_I) & \propto \int d z^1, \dots, d z^{M_P-1} d y_i \exp\left[-\frac{1}{2}(\vec{z} - \hat{C}_i\vec{m}_{P|I}(\vec{x}_I))^T(\hat{C}_i\hat\Sigma_P\hat{C}_i^T)^{-1}(\vec{z} - \hat{C}_i\vec{m}_{P|I}(\vec{x}_I))\right] \tanh y_i \\
    & \propto \int_{-\infty}^{+\infty} d y_i \exp\left[-\frac{1}{2 v_\mathrm{int}^i} \left(y_i - \frac{1}{M_P}\sum_{j = 1}^{M_P} A_{OP}^{ij }m_{P|I}^j(\vec{x}_I)\right)^2\right] \tanh y_i
\end{align*}
after the integration over $z^1, \dots, z^{M_P - 1}$. Here, we introduced the variance of the resulting marginal distribution
\begin{equation}
    v_\mathrm{int}^i = \sum_{j = 1}^{M_P} \sum_{k = 1}^{M_P} C_i^{M_Pj} \Sigma_P^{jk} C_i^{M_P k} = \frac{1}{M_P^2}\sum_{j = 1}^{M_P} \sum_{k = 1}^{M_P} A_{OP}^{ij} A_{OP}^{ik} \Sigma_P^{jk}
    \label{eqn:vINT}
\end{equation}
and mean
\begin{equation}
    \label{eqn:mINT}
    m_\mathrm{int}^i(\vec{x}_I) = \frac{1}{M_P} \sum_{j = 1}^{M_P} A_{OP}^{ij }m_{P|I}^j(\vec{x}_I) \;.
\end{equation}
Therefore, we can now proceed as in the previous case, so that
\begin{equation*}
    \ev{\phi_{OP}^i}^\mathrm{int}_P(\vec{x}_I) = \erf\left(\frac{m_\mathrm{int}^i(\vec{x}_I)}{\sqrt{2v_\mathrm{int}^i}}\right) + \sum_{n = 1}^\infty (-1)^n e^{2 n^2 v_\mathrm{int}^i} \left[V_n^+(m_\mathrm{int}^i(\vec{x}_I), v_\mathrm{int}^i) - V_n^-(m_\mathrm{int}^i(\vec{x}_I), v_\mathrm{int}^i)\right]
\end{equation*}
and the effective output operator now reads
\begin{align}
    \mathcal{L}_O^\mathrm{int,eff} (\vec{x}_I, \vec{x}_O) & = \sum_{i=1}^{M_O}\pdv{}{x_O^i} \left[\sum_{j=1}^{M_O}A_{O}^{ij}\left(x_O^j - m_{O|I}^{\mathrm{int}, j}(\vec{x}_I)\right) + D_O^i \pdv{}{x_O^i}\right]
\end{align}
where the mean is given by
\begin{align}
\label{eqn:SM:mOgI_FP_INT}
    m_{O|I}^{\mathrm{int}, i}(\vec{x}_I) = g_{OP} \sum_{j=1}^{M_O} \left(A_O^{-1}\right)^{ij} \Biggl[& \erf\left(\frac{m_\mathrm{int}^i(\vec{x}_I)}{\sqrt{2v_\mathrm{int}^i}}\right) + \nonumber \\
    & + \sum_{n = 1}^\infty (-1)^n e^{2 n^2 v_\mathrm{int}^i} \left[V_n^+\left(m_\mathrm{int}^i(\vec{x}_I), v_\mathrm{int}^i\right) - V_n^-(m_\mathrm{int}^i(\vec{x}_I), v_\mathrm{int}^i)\right] \Biggr] \; .
\end{align}
Notice that, as before, the mean of the output inherits a direct dependence on the input $\vec{x}_I$ through the averaging over the fast processing. However, contrary to the case of a nonlinear sum, the integration term makes the dependence on the inter-unit interactions highly nonlinear, as they appear in both $\vec{m}_\mathrm{int}$ and $\vec{v}_\mathrm{int}$.

\subsubsection{Computing the mutual information}
\noindent In both scenarios, we have an exact expression for $p_{O|I}^\mathrm{eff, st}$ which is a Gaussian distribution in the output state $\vec{x}_O$ with a nonlinear dependence on the input state $\vec{x}_I$. Therefore, we only need to solve the last order of the Fokker-Planck equation, which now reads
\begin{equation*}
    p_{P|I}^\mathrm{st} p_{O|I}^\mathrm{eff, st} \pdv{}{t}p_{IPO}^{(P, 0)} p_I^{(P, 0)} = p_{P|I}^\mathrm{st} p_{O|I}^\mathrm{eff, st} \mathcal{L}_I p_I^{(P, 0)} + \mathcal{L}_P p_{IPO}^{(O, 0)} 
\end{equation*}
at order $\mathcal{O}(1)$. By integrating over $\vec{x}_P$ and ignoring the vanishing border term, we end up with
\begin{equation}
    \pdv{}{t}p_I^{(P, 0)} = \mathcal{L}_I p_I^{(P, 0)}
\end{equation}
which immediately leads to the stationary Gaussian distribution $p_I^{(P, 0)} := p_I^\mathrm{st} = \mathcal{N}(0, \hat{\Sigma}_I)$, with $\hat{A}_I \hat{\Sigma}_I + \hat\Sigma_I\hat{A}_I^T = 2 \hat{D}_I$. This is expected, as the input evolves independently on the other degrees of freedom. Overall, we can write the leading-order solution of the joint probability distribution as follows:
\begin{equation}
    p_{IPO}^{(P, 0)} := p_{IPO}^\mathrm{fp} = p_{P|I}^\mathrm{st}\left(\vec{m}_{P|I}(\vec{x}_I)\right) p_{O|I}^\mathrm{eff, st}\left(\vec{m}_{O|I}(\vec{x}_I)\right)p_I^\mathrm{st}
\end{equation}
where the superscript stands for ``fast processing'' and we highlighted that the dependencies of the conditional distributions enter through their means. As in the case of two units, $p_{IPO}^\mathrm{fp}$ is a highly nonlinear distribution. However, our factorization into Gaussian distributions allows for its efficient sampling, as all the nonlinearities appear in the mean as conditional dependencies. In particular, we can compute the mutual information between the input and the output. We immediately have that their joint distribution is
\begin{equation*}
    p_{IO}^\mathrm{fp} = p_{O|I}^\mathrm{eff, st}p_I^\mathrm{st}
\end{equation*}
so that 
\begin{equation}
    h_{O|I}(\vec{x}_I) = \frac{1}{2}\left[M_O \left(1 + \log_2 (2\pi)\right) + \log_2 \det \hat\Sigma_O\right] \equiv h_{O|I}
\end{equation}
since the nonlinear dependencies of $p_{O|I}^\mathrm{eff, st}$ on $\vec{x}_I$ do not appear in its covariance matrix. Thus, we only need to evaluate $H_O$ numerically to estimate $I_{IO} = H_O - h_{O|I}$. We can proceed as detailed in Section \ref{sec:SM:IO}. We can easily sample the joint distribution $p_{IO}^\mathrm{fp}$ by leveraging its Gaussian factorization:
\begin{enumerate}
    \item sample $\{\vec{x}_I\}_{i = 1}^{N_\mathrm{sam}}$ from the independent Gaussian distribution of the input;
    \item compute the means $\vec{m}_{O|I}(\{\vec{x}_I\}_i)$ through either Eq.~\eqref{eqn:SM:mOgI_FP_NS} or Eq.~\eqref{eqn:SM:mOgI_FP_INT}, depending on the nonlinearity, for each sample $i$;
    \item for all $i$, sample $\vec{x}_O$ from the multivariate Gaussian with covariance $\hat{\Sigma}_O$ and means $\vec{m}_{O|I}(\{\vec{x}_I\}_i)$.
\end{enumerate}
Then, the entropy $H_O$ of the output distribution can be estimated from the samples $\{\vec{x}_O\}_i$ \cite{vasicek1976test, kozachenko1987sample}. Since we focus on one-dimensional outputs, such estimates are especially robust as they do not suffer from the curse of dimensionality.

\subsection{Slow processing}
\noindent We now focus on the case $\tau_I \gg \tau_P \gg \tau_O$, i.e., of a processing unit that is much slower than the output. As before, after rescaling time by the slowest timescale $t \to t/\tau_I$, we seek a solution of the form
\begin{equation*}
    p_{IPO}(\vec{x}_I, \vec{x}_P, \vec{x}_O, t) = p_{IPO}^{(O, 0)}(\vec{x}_I, \vec{x}_P, \vec{x}_O, t) + \epsilon_O\left[p_{IPO}^{(P, 0)}(\vec{x}_I, \vec{x}_P, \vec{x}_O, t) + \epsilon_P p_{IPO}^{(I, 1)}(\vec{x}_I, \vec{x}_P, \vec{x}_O, t) \right]  + \mathcal{O}\left(\epsilon_O^2, \epsilon_P^2\right)
\end{equation*}
where now $\epsilon_O = \tau_O/\tau_I \ll \epsilon_P = \tau_P/\tau_I$. At leading order, the Fokker-Planck equation becomes
\begin{equation*}
    \pdv{}{t}p_{IPO}^{(O, 0)} = \left[\mathcal{L}_I + \frac{\mathcal{L}_O}{\epsilon_O} + \frac{\mathcal{L}_P}{\epsilon_P}\right]p_{IPO}^{(O, 0)} + \mathcal{L}_P p_{IPO}^{(P, 0)}
\end{equation*}
and, at order $\mathcal{O}(1/\epsilon_O)$, we find that
\begin{equation*}
    \mathcal{L}_O (\vec{x}_P, \vec{x}_O) \, p_{IPO}^{(O, 0)}(\vec{x}_I, \vec{x}_P, \vec{x}_O, t) = 0 \;.
\end{equation*}
After a marginalization over the input state, $\vec{x}_I$, we have:
\begin{equation*}
    \mathcal{L}_O (\vec{x}_P, \vec{x}_O) \, p_{PO}^{(O, 0)}(\vec{x}_P, \vec{x}_O, t) = 0 = p_P^{(O,0)} (\vec{x}_P, t) \mathcal{L}_O (\vec{x}_P, \vec{x}_O) \, p_{O|P}^\mathrm{st}(\vec{x}_O | \vec{x}_P) \implies \mathcal{L}_{O|P} (\vec{x}_P, \vec{x}_O) \, p_{O|P}^\mathrm{st}(\vec{x}_O | \vec{x}_P) = 0
\end{equation*}
where for notational clarity $\mathcal{L}_{O|P} := \mathcal{L}_O$, denoting that its stationary distribution $p_{O|P}^\mathrm{st}$ is obtained at a fixed processing state. Thus, a solution of the form $p_{IPO}^{(P, 0)} = p_{O|P}^\mathrm{st} p_{IP}^{(O, 0)}$ automatically solves the leading order considered above. In particular, as in the previous section, we have that
\begin{equation}
    p_{O|P}^\mathrm{st}(\vec{x}_O | \vec{x}_P) = \mathcal{N}\left(\vec{m}_{O | P} (\vec{x}_P), \hat{\Sigma}_O\right)
\end{equation}
is a Gaussian distribution with a mean that depends nonlinearly on the processing state,
\begin{equation}
    m^i_{O|P}(\vec{x}_P) = g_{OP} \sum_{k=1}^{M_O}\left(A_O^{-1}\right)^{ik}\phi_{OP}\left(A_{OP}^{k, 1}, \dots, A_{OP}^{k, M_P}; \vec{x}_P\right), \quad i = 1, \dots, M_O,
\end{equation}
and a covariance obeying the Lyapunov equation $\hat{A}_O \hat{\Sigma}_O + \hat\Sigma_O\hat{A}_O^T = 2 \hat{D}_O$.

At the next order, $\mathcal{O}(1/\epsilon_P)$, we find
\begin{equation*}
    p_{O|P}^\mathrm{st}(\vec{x}_O | \vec{x}_P) \mathcal{L}_P(\vec{x}_I, \vec{x}_P) \, p_{IP}^{(O, 0)}(\vec{x}_I, \vec{x}_P, t) = 0 = p_{O|P}^\mathrm{st}(\vec{x}_O | \vec{x}_P) p_I^{(O,0)} (\vec{x}_I, t)  \mathcal{L}_P(\vec{x}_I, \vec{x}_P) \, p_{P|I}^{(O, 0), \mathrm{st}}(\vec{x}_P | \vec{x}_I) \; .
\end{equation*}
Contrarily to the case of the previous section, the integration over $\vec{x}_O$ can now be carried out immediately, since the operator does not depend on the output state, leading to
\begin{equation}
    \mathcal{L}_{P|I} (\vec{x}_I, \vec{x}_P) \, p_{P|I}^\mathrm{st}(\vec{x}_P | \vec{x}_I) = 0
\end{equation}
where, as before, we introduced $\mathcal{L}_{P|I} := \mathcal{L}_P$ to denote that its stationary distribution $p_{P|I}^\mathrm{st} = p_{P|I}^{(O, 0), \mathrm{st}}(\vec{x}_P | \vec{x}_I)$ is obtained at a fixed input state. In particular, this is once more a Gaussian distribution where the nonlinear dependencies enter in the form of conditional dependencies of the mean on $\vec{x}_I$:
\begin{equation}
    p_{P|I}^\mathrm{st}(\vec{x}_P | \vec{x}_I) = \mathcal{N}\left(\vec{m}_{P | I} (\vec{x}_I), \hat{\Sigma}_P\right)
\end{equation}
where
\begin{equation}
    m^i_{P|I}(\vec{x}_I) = g_{PI} \sum_{k=1}^{M_P}\left(A_P^{-1}\right)^{ik}\phi_{PI}\left(A_{PI}^{k, 1}, \dots, A_{PI}^{k, M_I}; \vec{x}_I\right), \quad i = 1, \dots, M_P,
\end{equation}
and the covariance matrix solves $\hat{A}_P \hat{\Sigma}_P + \hat\Sigma_P\hat{A}_P^T = 2 \hat{D}_P$. Thus, we end up with $p_{IPO}^{(O, 0)} = p_{O|P}^\mathrm{st} p_{P|I}^\mathrm{st}p_I^{(O, 0)}$.

At order $\mathcal{O}(1)$, finally, we simply have that 
\begin{equation}
    \pdv{}{t}p_I^{(O, 0)} = \mathcal{L}_I p_I^{(O, 0)} \implies p_I^{(O, 0)} := p_I^\mathrm{st} = \mathcal{N}(0, \hat{\Sigma}_I) 
\end{equation}
with $\hat{A}_I \hat{\Sigma}_I + \hat\Sigma_I\hat{A}_I^T = 2 \hat{D}_I$. Once more, this highlights that the input evolves independently on the other degrees of freedom. Overall, we find that the Fokker-Planck equation is solved at leading order by
\begin{equation}
    p_{IPO}^{(O, 0)} := p_{IPO}^\mathrm{sp} = p_{O|P}^\mathrm{st}\left(\vec{m}_{O|P}(\vec{x}_P)\right) p_{P|I}^\mathrm{st}\left(\vec{m}_{P|I}(\vec{x}_I)\right)p_I^\mathrm{st}
    \label{eqn:pslowproc}
\end{equation}
where the superscript stands for ``slow processing'' and we highlighted that the dependencies of the conditional distributions enter through their means. This expression is formally identical for both nonlinear scenarios, even if they change the internal functional dependencies and, as such, the overall shape of the distribution. Once again, $p_{IPO}^\mathrm{sp}$ is a highly nonlinear distribution, but the way the conditional dependencies appear is crucially different than the case of a fast processing unit and only depends on the timescale ordering considered. This is particularly relevant for the mutual information between the input and the output, since the distribution
\begin{equation}
    p_{IO}^\mathrm{sp}(\vec{x}_I, \vec{x}_O) = p_I^\mathrm{st}(\vec{x}_I) p^\mathrm{st}_{O|I}(\vec{x}_O | \vec{x}_I) = p_I^\mathrm{st}(\vec{x}_I)\int d\vec{x}_P p_{O|P}^\mathrm{st}(\vec{x}_O | \vec{x}_P) p_{P|I}^\mathrm{st}(\vec{x}_P | \vec{x}_I)
\end{equation}
cannot be easily computed. Thus, the entropy of the conditional distribution $h_{O|I}(\vec{x}_I)$ is not known analytically.

To address this issue, we exploit the fact that we can efficiently sample $p^\mathrm{st}_{O|I}$, allowing us to easily obtain a numerical estimate of $h_{O|I}(\vec{x}_I)$. Then, we can estimate the conditional entropy
\begin{equation}
    H_{O|I} = \int d\vec{x}_I p_I^\mathrm{st}(\vec{x}_I)h_{O|I}(\vec{x}_I)
\end{equation}
with importance sampling. We proceed as follows:
\begin{enumerate}
    \item sample a fixed input $\vec{x}_I^{(i)} \sim \mathcal{N}(0, \hat{\Sigma}_I)$ for $i = 1, \dots, N_{\mathrm{sam},I}$;
    \item for each input sample $\vec{x}_I^{(i)}$, compute $\vec{m}_{P|I}\left(\vec{x}_I^{(i)}\right)$, and extract the samples $\vec{x}_P^{(i, j)}$ from $\mathcal{N}\left(\vec{m}_{P|I}\left(\vec{x}_I^{(i)}\right), \hat{\Sigma}_P\right)$ for $j = 1, \dots, N_\mathrm{sam}$;
    \item for each processing sample $\vec{x}_P^{(i, j)}$, compute the mean $\vec{m}_{O|P}\left(\vec{x}_P^{(i, j)}\right)$ and extract the corresponding output $\vec{x}_O^{(i, j)}$ from $\mathcal{N}\left(\vec{m}_{O|P}\left(\vec{x}_P^{(i, j)}\right), \hat{\Sigma}_O\right)$;
    \item for each input sample $\vec{x}_I^{(i)}$, estimate the entropy $h_{O|I}\left(\vec{x}_I^{(i)}\right)$ of the conditional distribution $p_{O|I}^\mathrm{st}$ from the output samples $\{\vec{x}_O\}_{i,j}$, using any numerical estimator (e.g., Vasicek \cite{vasicek1976test} or Kozachenko-Leonenko \cite{kozachenko1987sample});
    \item estimate the conditional entropy $H_{O|I}$ via importance sampling,
    \begin{equation}
        H_{O|I} \approx \sum_{i = 1}^{N_{\mathrm{sam}, I}} h_{O|I}\left(\vec{x}_I^{(i)}\right)
    \end{equation}
    \item from all the output samples $\{\vec{x}_O\}_{i,j}$, estimate the entropy $H_O$ via any numerical estimator;
    \item compute the mutual information as $I_{IO} = H_O - H_{O|I}$.
\end{enumerate}
Once more, since we focus on one-dimensional outputs, this sampling scheme avoids any issue with the curse of dimensionality, allowing us to explore large processing (and input) dimensions.

\begin{figure}
    \centering
    \includegraphics[width=\textwidth]{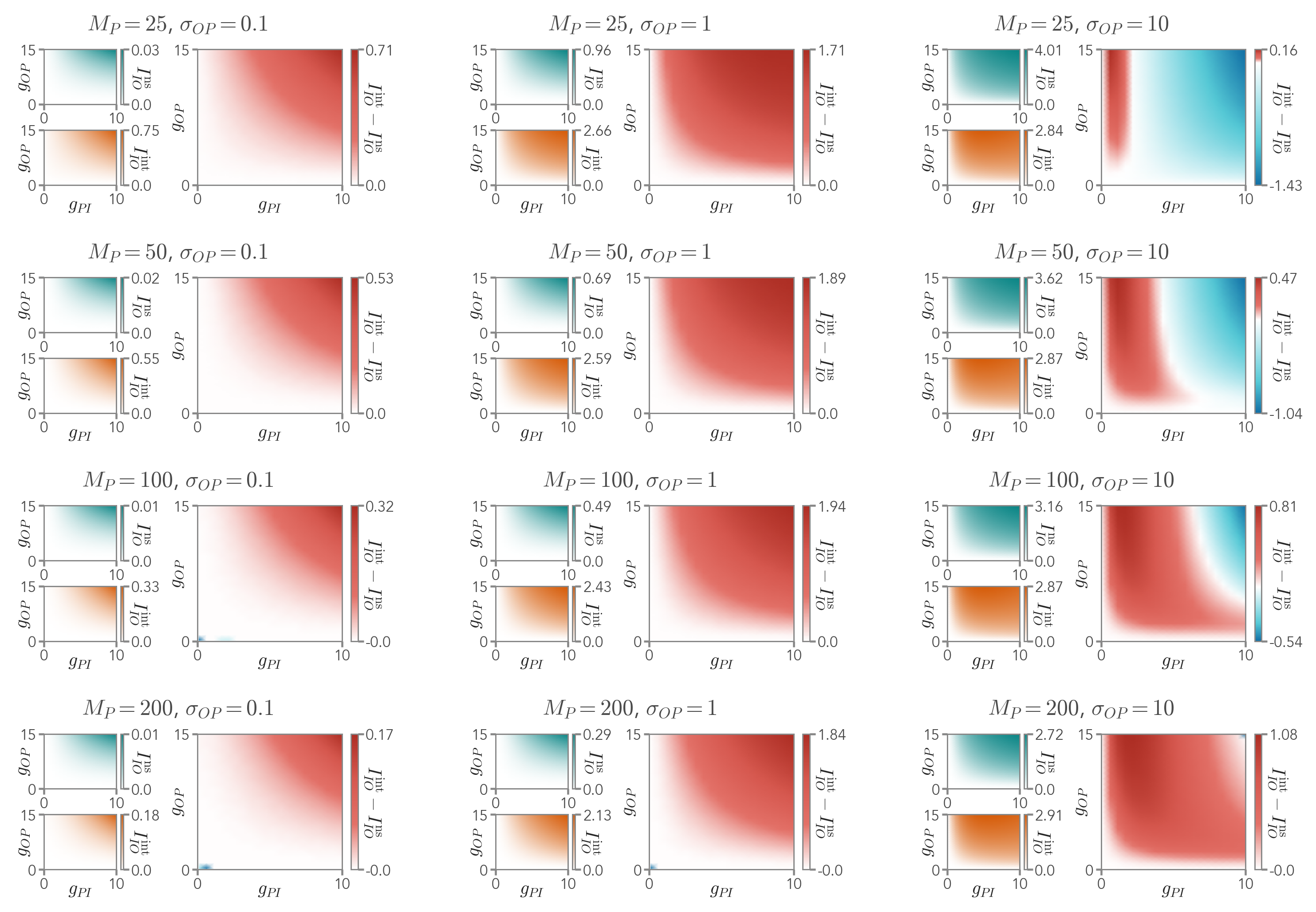}
    \caption{Mutual information between the input and the output in the fast processing case, both for an activation function implementing a nonlinear summation ($I_{IO}^\mathrm{ns}$, teal) and a nonlinear integration ($I_{IO}^\mathrm{int}$, orange), for different size of the processing unit ($M_P$) and with different standard deviations $\sigma_{OP}$ of the interaction matrix $\hat{A}_{OP} \sim \mathcal{N}(0, \sigma_{OP})$. In this plot, $M_I = 50$, $\sigma_{II} = \sigma_{PP} = 0.9$, and $\sigma_{PI} = 1$. At large variances and in strong coupling regimes, a high-dimensional processing unit is needed for nonlinear integration to provide more information than nonlinear summation. This is due to the fact that, if $M_p$ is small and $\sigma_{OP}$ is large, the entries of the interaction matrix $\hat{A}_{OP}$ will be very different, pushing the activation function in the saturation regime. All information is measured in bits. Results are averaged over $10^3$ realization of the random matrices. For each realization, $N_\mathrm{sam} = 10^4$.}
    \label{fig:SM:phase_space_mPP}
\end{figure}

\subsection{Effect of the dimensionality of the processing}
\noindent We now briefly study the interplay between the dimensionality of the processing unit, $M_P$, and the couplings between the units. In particular, as in the main text, we take the internal couplings of the units to be described by the random matrices
\begin{equation}
    A_I^{ij} \sim \mathcal{N}\left(0, \frac{\sigma_{II}}{\sqrt{M_I}}\right), \quad A_P^{ij} \sim \mathcal{N}\left(0, \frac{\sigma_{II}}{\sqrt{M_I}}\right), \quad A_O = 1
\end{equation}
where the output is one-dimensional, and the usual normalization of the standard deviations of the elements of $\hat{A}_I$ and $\hat{A}_P$ ensures a proper scaling of their spectrum with the dimensionality of the units. Similarly, the interactions between the units are given by the random matrices
\begin{equation}
    A_{PI}^{ij} \sim \mathcal{N}\left(0, \sigma_{PI}\right), \quad A_{OP}^{ij} \sim \mathcal{N}\left(0, \sigma_{OP}\right)
\end{equation}
and we fix $\sigma_{OP} = 1$ for simplicity.

\begin{figure}
    \centering
    \includegraphics[width=\textwidth]{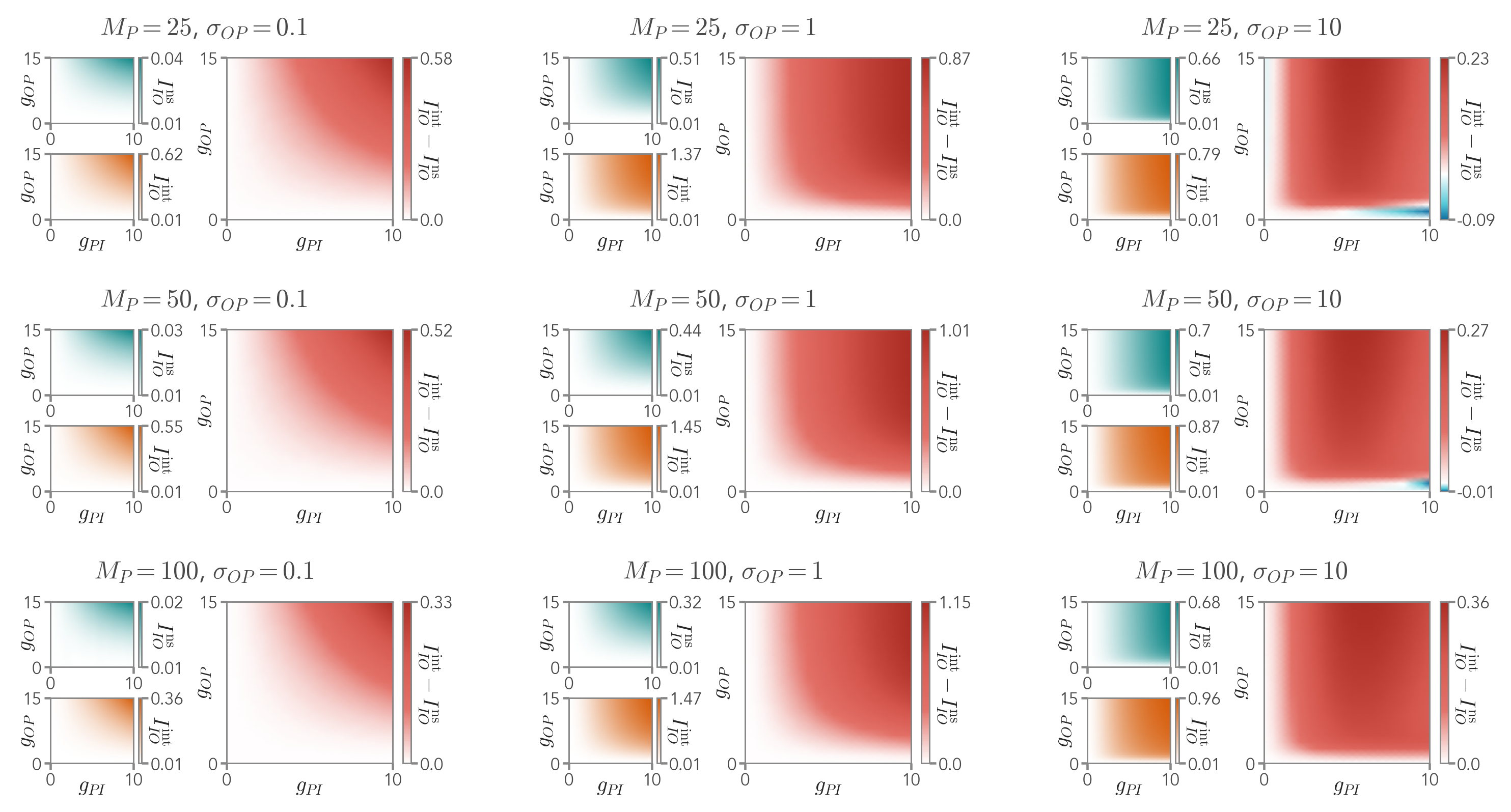}
    \caption{Mutual information between the input and the output in the slow processing case, both for an activation function implementing a nonlinear summation ($I_{IO}^\mathrm{ns}$, teal) and a nonlinear integration ($I_{IO}^\mathrm{int}$, orange), for different size of the processing unit ($M_P$) and with different standard deviations $\sigma_{OP}$ of the interaction matrix $\hat{A}_{OP} \sim \mathcal{N}(0, \sigma_{OP})$. In this plot, $M_I = 50$, $\sigma_{II} = \sigma_{PP} = 0.9$, and $\sigma_{PI} = 1$. In this slow-processing case, information is typically smaller than in the fast-processing one. However, information is larger with nonlinear integration rather than nonlinear summation even for large variances $\sigma_{OP}^2$ and small processing dimensions $M_P$. All information is measured in bits. Results are averaged over $10^3$ realization of the random matrices. For each realization, $N_{\mathrm{sam}, I} = 2\cdot10^3$ and $N_{\mathrm{sam}} = \cdot10^3$.}
    \label{fig:SM:phase_space_FF}
\end{figure}

We first consider the case of a fast processing unit. In Figure \ref{fig:SM:phase_space_mPP}, we show how the mutual information between the input and the output behaves as a function of the couplings $g_{PI}$ and $g_{OP}$ and at different values of $M_P$ and $\sigma_{OP}$, for an activation function implementing both a nonlinear summation and a nonlinear integration. Remarkably, we find that the dimensionality of the processing unit has a relevant effect at large interaction variances $\sigma_{OP}^2$. In particular, if $M_P$ is small enough and in strong coupling regimes, we find that the nonlinear summation may provide a larger mutual information than the nonlinear integration. Intuitively, this happens due to the highly nonlinear dependencies of the mean $\vec{m}_{O|I}$ (Eqs.~\eqref{eqn:SM:mOgI_FP_NS} and \eqref{eqn:SM:mOgI_FP_INT}) on interaction matrix $\hat{A}_{OP}$ and the variance of the specific realization of its elements. As $M_P$ increases, the elements of $\hat{A}_{OP}$ are more uniformly sampled from the underlying Gaussian distribution, and this effect becomes less and less prominent and eventually disappears. Furthermore, we find that at small variances ($\sigma_{OP} = 0.1$) the mutual information $I_{IO}$ tends to be vanishingly small for a nonlinear summation, whereas it is significantly larger in the nonlinear integration case, especially at low processing dimensionalities. 

Then, we switch to the case of a slow processing unit in Figure \ref{fig:SM:phase_space_FF}. Remarkably, the dimensionality of the processing is less impactful. In particular, in the coupling regimes we explored, the region where $I_{IO}^\mathrm{ns} > I_{IO}^\mathrm{int}$ is smaller and becomes negligible already at $M_P = 50$. Furthermore, we find that the input-output mutual information is consistently smaller with respect to the fast-processing case, as in the main text, suggesting that the timescales of the different units play a quantitative role in determining the information-processing capabilities of the system.

\begin{figure}
    \centering
    \includegraphics[width=0.8\textwidth]{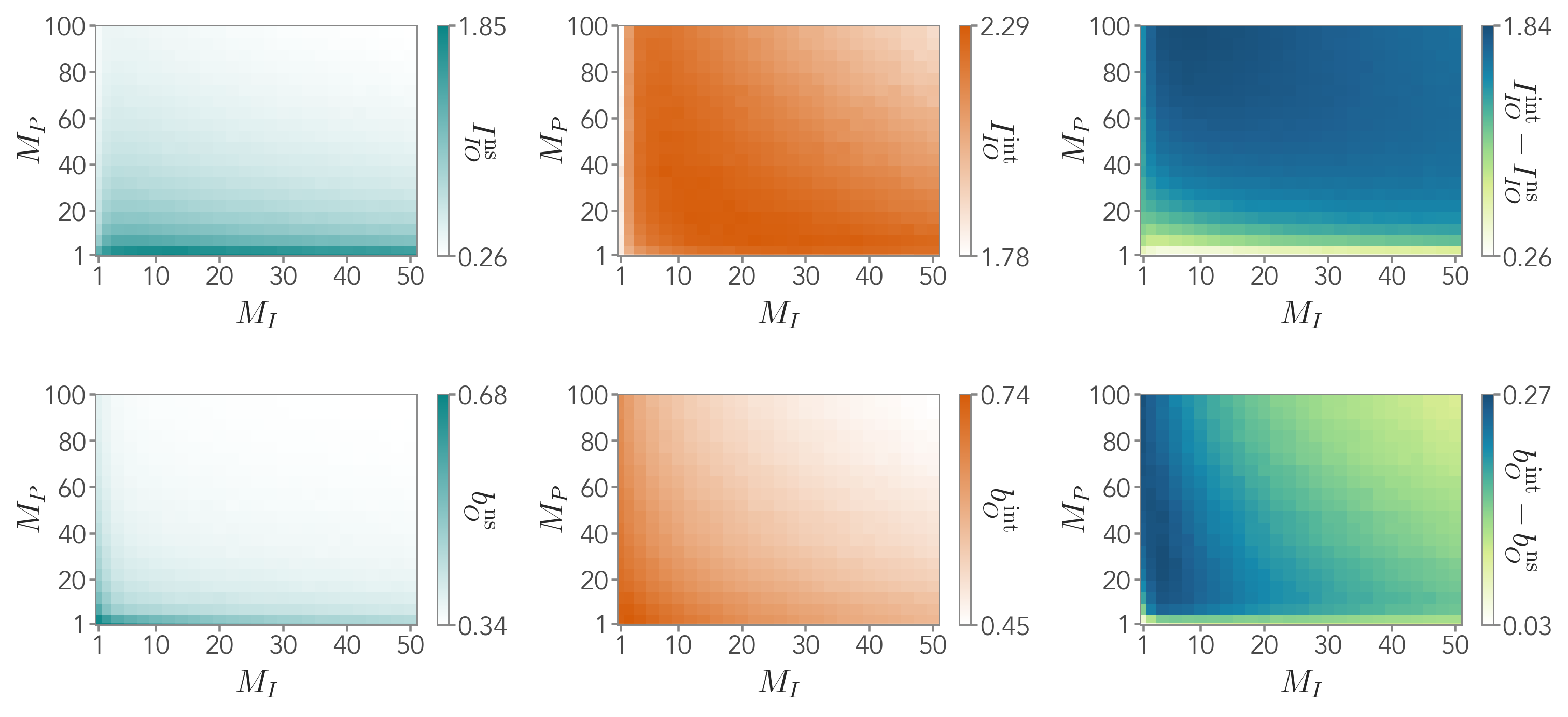}
    \caption{Mutual information between the input and the output (first row) and bimodality coefficient of the output distribution (second row) for a system with a fast processing unit, with $g_{PI} = g_{OP} = 10$, $\sigma_{OP} = \sigma_{PI} = 1$, and $\sigma_{II} = \sigma_{PP} = 0.9$. For nonlinear integration (orange), information is higher at large $M_P$ for small input dimensions, and vice-versa. For nonlinear summation (teal), instead, information tends to decrease with the processing dimensionality. Both types of activation functions may display a bistable output distribution, particularly for smaller dimensions, although nonlinear integration enhances bimodality. All information is measured in bits. Results are averaged over $10^3$ realization of the random matrices. For each realization, $N_{\mathrm{sam}} = 2\cdot 10^3$.}
    \label{fig:SM:phase_space_MIxMP}
\end{figure}

Finally, in Figure \ref{fig:SM:phase_space_MIxMP} we show how $I_{IO}$ changes with $M_I$ and $M_P$ for a fast processing unit. We find that information typically decreases with $M_P$, regardless of the input dimension, when the activation function implements a nonlinear summation. However, the picture is markedly different when we switch to nonlinear integration, as reported in the main text. At small input dimensions, the mutual information between the input and the output is higher for large processing dimensionalities, suggesting that a nonlinear embedding of a low-dimensional input in a higher-dimensional space favors information processing. On the contrary, $I_{IO}$ is maximal at small $M_P$ for large $M_I$, so that information processing is favored by a nonlinear compression of the input in a lower-dimensional processing space. 

\section{Emergent output bistability}
\noindent As shown in the main text, the output distribution may be bimodal, depending on the choice of the processing parameters. We measure this bistability of the underlying Langevin dynamics $x_O(t)$ by computing Sarle's bimodality coefficient \cite{pfister2013good}, defined as:
\begin{equation}
    b = \frac{s^2 + 1}{\kappa + q(n_\mathrm{samples})}
\end{equation}
where $n_\mathrm{samples}$ is the number of samples at hand, $q(x) = 3(n-1)^2/[(n-2)(n-3)]$, $s$ is the sample skewness,
\begin{equation}
    s = \frac{1}{n_\mathrm{samples}} \frac{\sum_{i} \left[x_O^{(i)} - \ev{x_O}\right]^3}{\left[\ev{x_O^2} - \ev{x_O}^2\right]^{3/2}}
\end{equation}
and $\kappa$ is the excess kurtosis,
\begin{equation}
    \kappa = \frac{1}{n_\mathrm{samples}} \frac{\sum_{i} \left[x_O^{(i)} - \ev{x_O}\right]^4}{\left[\ev{x_O^2} - \ev{x_O}^2\right]^{2}} - 3 \; .
\end{equation}
We note that the bimodality coefficient can take values between $0$ and $1$, with $b = 1$ for a perfectly bimodal distribution such as the sum of two Dirac's delta functions centered at different points. It is also easy to check that, for a uniform distribution, we have $b = 5/9 \approx 0.55$. Hence, a value significantly higher than this may indicate a high degree of bistability. In Figure \ref{fig:SM:phase_space_MIxMP}, we show how the bimodality coefficient of the output distribution behaves as a function of $M_I$ and $M_P$ in a system with a fast processing unit, as in the main text but for a broader parameter range. We find that both for nonlinear summation and integration smaller dimensions favor the bistability, as the small sizes of the random matrices at play favor the presence of more diverse elements. Crucially, nonlinear integration typically features a higher bimodality coefficient, suggesting that the sensitivity of the output distribution is enhanced by integration.  

\subsection{Tuning the output bistability}
\noindent So far, we have considered the nonlinearity implemented through a hyperbolic tangent centered in zero, i.e., with a functional form of the type $\tanh{x}$, in the two different settings of nonlinear summation and integration. To gain more insights into the origin of the bistable dynamics observed in the main text, here we add an internal parameter to the hyperbolic tangent shaping the interaction between the processing and the output unit. That is, we add a parameter $\theta$ tuning the saturation regimes by modulating its argument, i.e., $x \to x - \theta$. We detail how to incorporate this parameter into our calculations, leading to the fact that, as shown in the main text, the net effect of the presence of this additional parameter is to tune the height of the distribution peaks in the regimes in which the output presents an emergent bimodal distribution.

The case of slow processing does not require additional analysis, as the structure of the joint distribution is valid for any nonlinear mechanism (see Eq.~\eqref{eqn:pslowproc}). As such, the presence of additional parameters will not change this result. On the other hand, the solution in the case of fast processing is sensitive to the form of the nonlinearity under consideration due to its Gaussian average appearing in the effective distributions. Let us start with the nonlinear sum. By following the steps highlighted above, obtaining a solution for this scenario amounts to determining the effective Fokker-Planck operator, $\mathcal{L_{O|I}^{\rm NS, \rm eff}}$, whose drift depends on
\begin{align*}
    \ev{\phi_{OP}^i}_P^\mathrm{ns}(\vec{x}_I) & \propto \int d x_P^1 \dots d x_P^j \dots d x_P^{M_P} \exp\left[-\frac{1}{2}(\vec{x}_P - \vec{m}_{P|I}(\vec{x}_I))^T\hat\Sigma_P^{-1}(\vec{x}_P - \vec{m}_{P|I}(\vec{x}_I))\right]  \sum_{j = 1}^{M_P} \frac{A_{OP}^{ij}}{M_P}\tanh{(x_P^j - \theta_P^j)} \\
    & \propto \sum_{j = 1}^{M_P} \frac{A_{OP}^{ij}}{M_P} \int d x_P^j \exp\left[-\frac{1}{2 \Sigma_P^{jj}} \left(x_P^j - m_{P|I}^j(\vec{x}_I)\right)^2\right] \tanh{(x_P^j - \theta_P^j)}
\end{align*}
where we neglected the normalization terms for brevity and, as before, exploited the fact that the hyperbolic tangent term depends on the components of $\vec{x}_P$ separately. Moreover, we allowed for the presence of as many internal parameters $\theta_P^j$ as the number of processing states, for the sake of generality. For each $j$, this integral is of the form
\begin{align*}
    \int_{-\infty}^{+\infty} dz \tanh(z-\theta) e^{-\frac{(z-m)^2}{2\sigma^2}} \xrightarrow{\zeta = z - \theta} \left[\int_{0}^{+\infty} d\zeta \tanh(\zeta) e^{-\frac{(\zeta-(m-\theta))^2}{2\sigma^2}} - \int_0^{+\infty} d\zeta \tanh(\zeta) e^{-\frac{(\zeta+(m-\theta))^2}{2\sigma^2}}\right] \;.
\end{align*}
It can be solved following the procedure outlined above and the result in Eq.~\eqref{eqn:SM:apptanh} will only present an average shifted by $\theta$. Putting all the elements together, we obtained the following drift of the effective operator:
\begin{align*}
    m_{O|I}^{\mathrm{ns}, i}(\vec{x}_I) = \frac{g_{OP}}{M_P} \sum_{j=1}^{M_I} \sum_{k = 1}^{M_P}\left(A_O^{-1}\right)^{ij}A_{OP}^{jk} \Biggl[& \erf\left(\frac{m_{P|I}^k(\vec{x}_I) - \theta_P^k}{\sqrt{2\Sigma_P^{kk}}}\right) + \nonumber \\
    & + \sum_{n = 1}^\infty (-1)^n e^{2 n^2 \Sigma_P^{kk}} \left[V_n^+\left(m_{P|I}^k(\vec{x}_I) - \theta_P^k, \Sigma_P^{kk}\right) - V_n^-(m_{P|I}^k(\vec{x}_I) - \theta_P^k, \Sigma_P^{kk})\right] \Biggr] \; .
\end{align*}
We now move to the case of nonlinear integration. In this scenario, the effective Fokker-Planck operator $\mathcal{L_{O|I}^{\rm INT, \rm eff}}$ depends on the following integral:
\begin{align*}
    \ev{\phi_{OP}^i}^\mathrm{int}_P(\vec{x}_I) & \propto \int d x_P^1 \dots d x_P^{M_P} \exp\left[-\frac{1}{2}(\vec{x}_P - \vec{m}_{P|I}(\vec{x}_I))^T\hat\Sigma_P^{-1}(\vec{x}_P - \vec{m}_{P|I}(\vec{x}_I))\right] \tanh\left(\sum_{j = 1}^{M_P} \frac{A_{OP}^{ij}}{M_P}(x_P^j - \theta_P^j)\right) \; .
\end{align*}
By performing the same change of variable employed above and determined by the matrix in Eq.~\eqref{eqn:changevariable}, we obtain:
\begin{align*}
    \ev{\phi_{OP}^i}^\mathrm{int}_P(\vec{x}_I) & \propto \int d z^1, \dots, d z^{M_P-1} d y_i \exp\left[-\frac{1}{2}(\vec{z} - \hat{C}_i\vec{m}_{P|I}(\vec{x}_I))^T(\hat{C}_i\hat\Sigma_P\hat{C}_i^T)^{-1}(\vec{z} - \hat{C}_i\vec{m}_{P|I}(\vec{x}_I))\right] \tanh{(y_i - \Theta_i)} \\
    & \propto \int_{-\infty}^{+\infty} d y_i \exp\left[-\frac{1}{2 v_\mathrm{int}^i} \left(y_i - \frac{1}{M_P}\sum_{j = 1}^{M_P} A_{OP}^{ij }m_{P|I}^j(\vec{x}_I)\right)^2\right] \tanh{(y_i - \Theta_i)}
\end{align*}
where $\Theta_i = \sum_{j=1}^{M_P} \theta_P^j (A^{ij}_{OP}/M_P)$ is a global tuning parameter that depends on all output-processing interactions. As above, this integral can be carried out by exploiting the expansion of the hyperbolic tangent leading to:
\begin{align*}
    m_{O|I}^{\mathrm{int}, i}(\vec{x}_I) = g_{OP} \sum_{j=1}^{M_I} \left(A_O^{-1}\right)^{ij} \Biggl[& \erf\left(\frac{m_\mathrm{int}^i(\vec{x}_I) - \Theta_i}{\sqrt{2v_\mathrm{int}^i}}\right) + \nonumber \\
    & + \sum_{n = 1}^\infty (-1)^n e^{2 n^2 v_\mathrm{int}^i} \left[V_n^+\left(m_\mathrm{int}^i(\vec{x}_I) - \Theta_i, v_\mathrm{int}^i\right) - V_n^-(m_\mathrm{int}^i(\vec{x}_I) - \Theta_i, v_\mathrm{int}^i)\right] \Biggr]
\end{align*}
where $v_\mathrm{int}^i$ and $m_\mathrm{int}^i$ have been defined in Eqs.~\eqref{eqn:vINT} and \eqref{eqn:mINT}.

\end{document}